%% file: paper.tex
\documentclass[cernpreprint, UKenglish,texlive=2011,txfonts,texmf]{atlasdoc}

\pdfoutput=1

\usepackage{atlaspackage}

\usepackage{atlasbiblatex}

\usepackage{atlascontribute}

\usepackage{atlasphysics}

\usepackage{atlascover}
\newcommand{\mt}{\mbox{$m_\mathrm{T}^W$}} 
\newcommand{\MQ}{\mbox{$m_{Q}$}} 
\newcommand{\Mreco}{\mbox{$m_\text{reco}$}}
\newcommand{\Whad}{\mbox{$W_\text{had}$}}
\newcommand{\Wlep}{\mbox{$W_\text{lep}$}}

\usepackage{fancyvrb}

\usepackage{amsmath}
\usepackage{multirow}
\usepackage{bm}
\usepackage{graphicx} 
\usepackage{grffile}
\usepackage{float} 
\floatstyle{plaintop}
\restylefloat{table}
\usepackage{rotating}
\usepackage{placeins}
\usepackage{hyperref} 
\usepackage{multicol}

\AtlasTitle{Search for pair production of a new heavy quark that decays into a $W$ boson and a light quark in $pp$ collisions at $\sqrt{s} = 8$\,TeV with the ATLAS detector}
\author{The ATLAS Collaboration}

\date{\today}

\AtlasRefCode{EXOT-2014-10}

 \PreprintIdNumber{CERN-PH-EP-2015-212}

\AtlasDate{\today}

\AtlasJournal{PRD}

\AtlasAbstract{%
A search is presented for pair production of a new heavy quark ($Q$) that decays into a $W$ boson and a light quark ($q$) in the final state where one $W$ boson decays leptonically (to an electron or muon plus a neutrino) and the other $W$ boson decays hadronically. The analysis is performed using an integrated luminosity of 20.3\,fb$^{-1}$ of $pp$ collisions at $\sqrt{s} = 8\,$TeV collected by the ATLAS detector at the LHC.  No evidence of $Q\bar{Q}$ production is observed.  New chiral quarks with masses below 690\,GeV are excluded at 95\% confidence level, assuming BR$(Q\to Wq)=1$.  Results are also interpreted in the context of vectorlike quark models, resulting in the limits on the mass of a vectorlike quark in the two-dimensional plane of BR$(Q\to Wq)$ versus BR$(Q\to Hq)$.}

\hypersetup{pdftitle={ATLAS draft},pdfauthor={The ATLAS Collaboration}}

\begin{document}

\maketitle

\clearpage

\section{Introduction}
\label{sec:Introduction}

The recent observation of the long-predicted Higgs boson by the ATLAS
\cite{Aad:2012tfa} and CMS \cite{Chatrchyan:2012ufa} collaborations
now completes the Standard Model (SM).  In spite of its success, the
SM cannot account for dark matter and the matter/antimatter asymmetry
in the Universe and also fails to provide insight into the fermion
mass spectrum, nonzero neutrino masses, why there are three
generations of fermions, or why parity is not violated in the strong
interaction.  Furthermore, the observed Higgs boson is unnaturally
light, requiring fine tuning to cancel radiative corrections that
would naturally result in a mass many orders of magnitude larger, a
discrepancy known as the hierarchy problem~\cite{Susskind:1978ms}.

A variety of models have been proposed to address the various
shortcomings of the SM.  For example, a primary motivation for
supersymmetry (SUSY) is to solve the hierarchy
problem~\cite{Dimopoulos:1981zb}.  In SUSY models, the quadratically
divergent radiative corrections to the Higgs-boson mass due to SM
particles are automatically canceled by the corrections from the
supersymmetric partners.
Models such as Little Higgs, Composite Higgs, and Topcolor take a
different approach, proposing that electroweak symmetry breaking
happens dynamically as the result of a new strong
interaction~\cite{ArkaniHamed:2002qy,Contino,Dobrescu:1997nm,
  Hill:1994hp,Hill2003235}.  Grand unified theories (GUTs) go further,
unifying the electroweak and strong forces by proposing that the SM
gauge symmetry $S\!U(3)_\text{C} \times S\!U(2)_\text{L} \times
U(1)_\text{Y}$ is the low-energy limit of a single fundamental
symmetry group such as $S\!O(10)$ or
$E_6$~\cite{Kang:2007ib,Sultansoy}, which could potentially explain
the observed spectrum of fermions and even provide insight into the
unification of the electroweak and strong forces with gravity.  A
feature in many of these, and other
models~\cite{PhysRevD.41.1286,other1,other2}, is the prediction of
vectorlike quarks (VLQs), hypothetical spin-$\frac{1}{2}$ particles
that are triplets under the color gauge group and have identical
transformation properties for both chiralities under the electroweak
symmetry group. Furthermore, massive VLQs would respect gauge
invariance without coupling to the Higgs field.
This allows VLQs to avoid constraints from Higgs-boson production~\cite{nochiral}; 
if the Higgs sector is minimal, these constraints rule out additional chiral quarks.
However, some two-Higgs-doublet models are able to avoid those constraints and
accommodate a fourth generation of quarks~\cite{BarShalom:2012ms}.

In the models of interest, the VLQs have some mixing with the SM
quarks, allowing them to decay to SM quarks and either a $W$, $Z$, or
Higgs boson; however, the exact nature of the coupling depends on the
model.  For example, in composite Higgs models, the VLQs are involved
in a seesaw mechanism with the SM quarks, so the lightest VLQ couples
almost exclusively to the heaviest SM quarks ($t$- and
$b$-quarks)~\cite{Contino}.  However, there are also models that
predict \TeV-scale VLQs that could preferentially decay to light SM
quarks ($q = u, d, s,$ or
$c$)~\cite{Kang:2007ib,Sultansoy,Chakdar:2013tca}.  For example, the
left-right mirror model (LRMM)~\cite{Chakdar:2013tca} predicts three
generations of heavy ``mirror'' quarks, with the lightest mirror
generation coupling to the lightest SM generation.  The two lightest
mirror quarks could be pair-produced at the LHC via the strong
interaction and would then decay to $Wq$, $Zq$, or $Hq$ ($q=u$ or
$d$).  The LRMM would provide an explanation for tiny neutrino masses,
parity violation in weak interactions, parity conservation in strong
interactions, and could be the first step toward uncovering the
symmetry structure of a GUT.  Another model predicting VLQs that decay
to light quarks is the $E_6$ GUT with isosinglet
quarks~\cite{Kang:2007ib,Sultansoy}.  In this model, after the $E_6$
symmetry is broken down to the SM group structure, VLQ partners to the
$d$-, $s$-, and $b$-quarks are predicted.  If the VLQs have the same
mass ordering as their SM partners, the lightest VLQ would couple
predominately to first-generation SM quarks ($q = u$ or $d$).  The
values for the branching ratios to the three decay modes ($Wq$, $Zq$,
$Hq$) depend on parameters in the model.  The values in the $E_6$
isosinglet model range from approximately (0.6, 0.3, 0.1) to (0.5,
0.25, 0.25), while the LRMM allows branching ratios from approximately
(0.6, 0.4, 0) to (0, 0, 1), depending on the VLQ mass and mixing
angles.

If such new quarks exist, they are expected to be produced predominantly in pairs via
the strong interaction for masses up to $O(1\TeV$) in LHC collisions with a center-of-mass energy of 8~\TeV.  
Single production of a new heavy quark, $Q$, would be dominant for very
high quark masses, but the production rate is model-dependent and
could be suppressed if the coupling to SM quarks is small.
To date, there have been two analyses of LHC data sensitive to VLQs that decay
to light quarks, both using 1.04\,\ifb{} of ATLAS data at $\sqrt{s} =
7\TeV$: one search for pair-production~\cite{Aad:2012bt} and one for
single production~\cite{Aad:2011yn}.  The pair-production search set a
lower limit on the $Q$ mass of 350~\GeV{} at 95\% confidence level,
assuming BR($Q\to Wq$) = 1.  Such a signal was also ruled out by the
Tevatron for masses up to 340~\GeV{}~\cite{Aaltonen:2011tq}.

\begin{figure}[htbp]
\begin{center}
\includegraphics[scale=0.28]{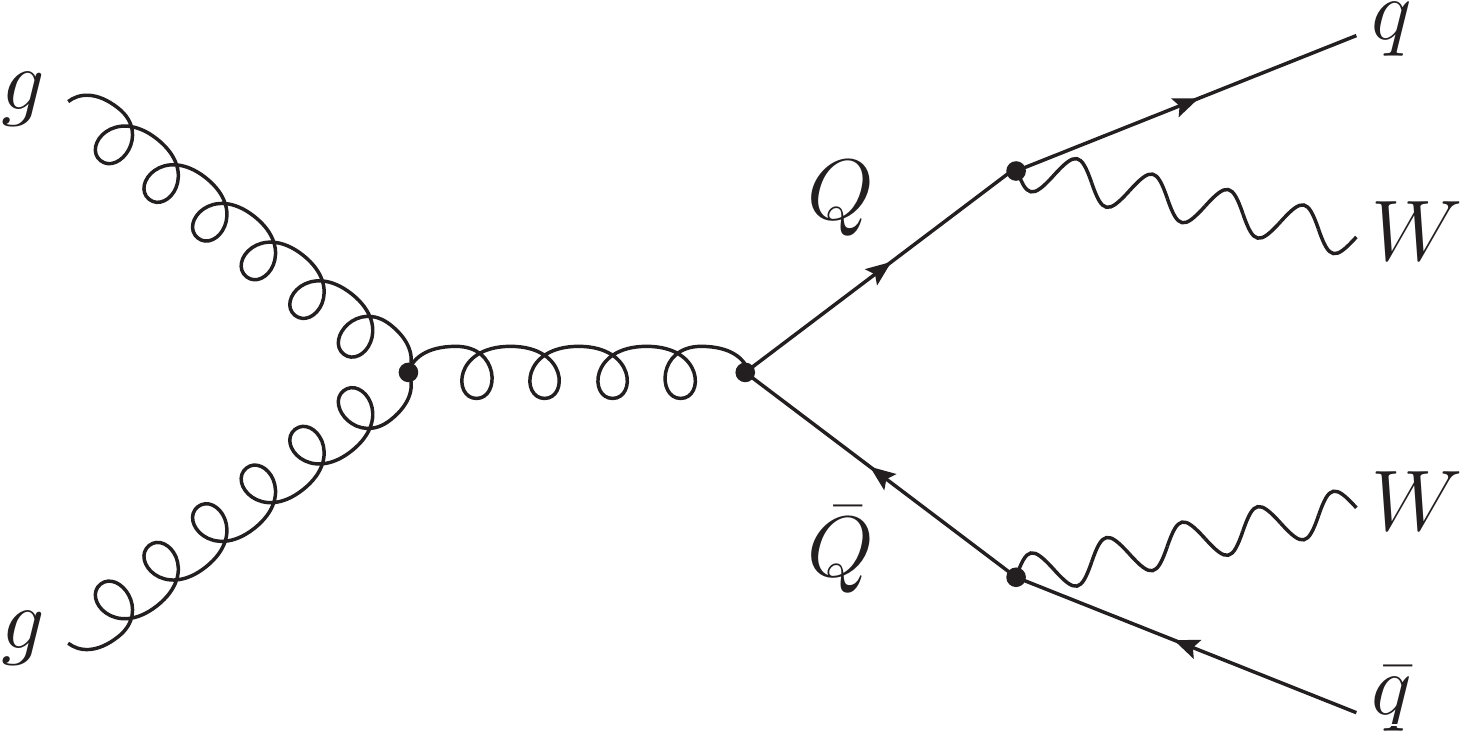}
\hspace{2mm}
\includegraphics[scale=0.28]{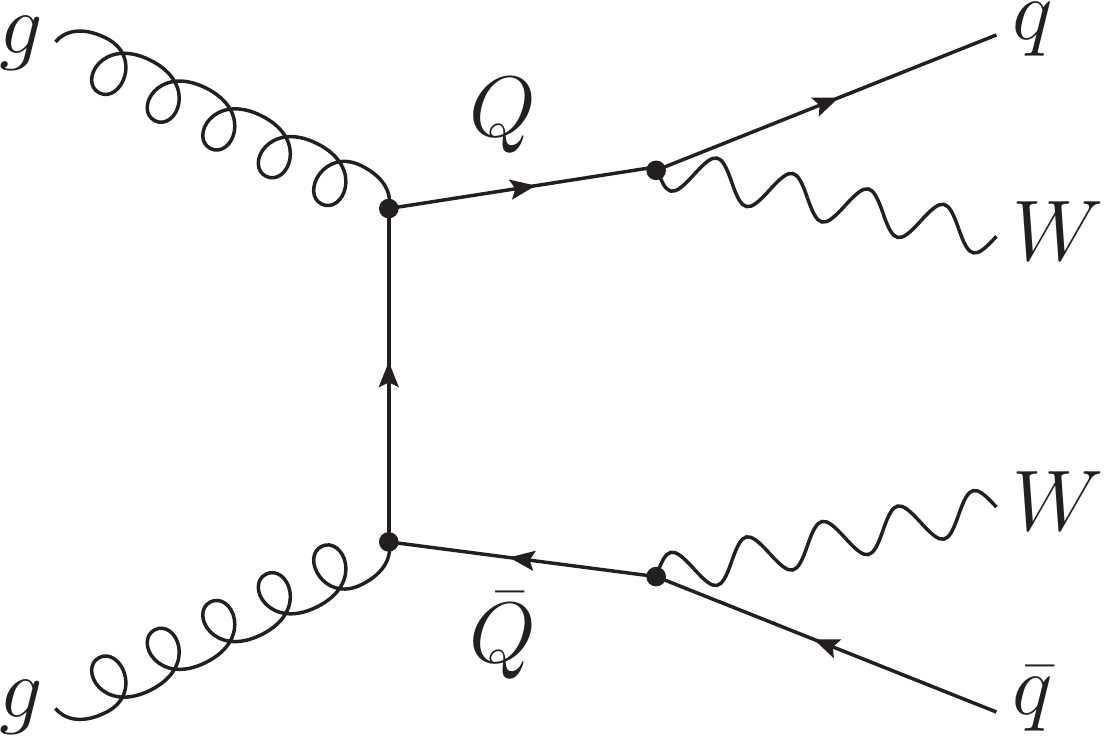}
\hspace{2mm}
\includegraphics[scale=0.28]{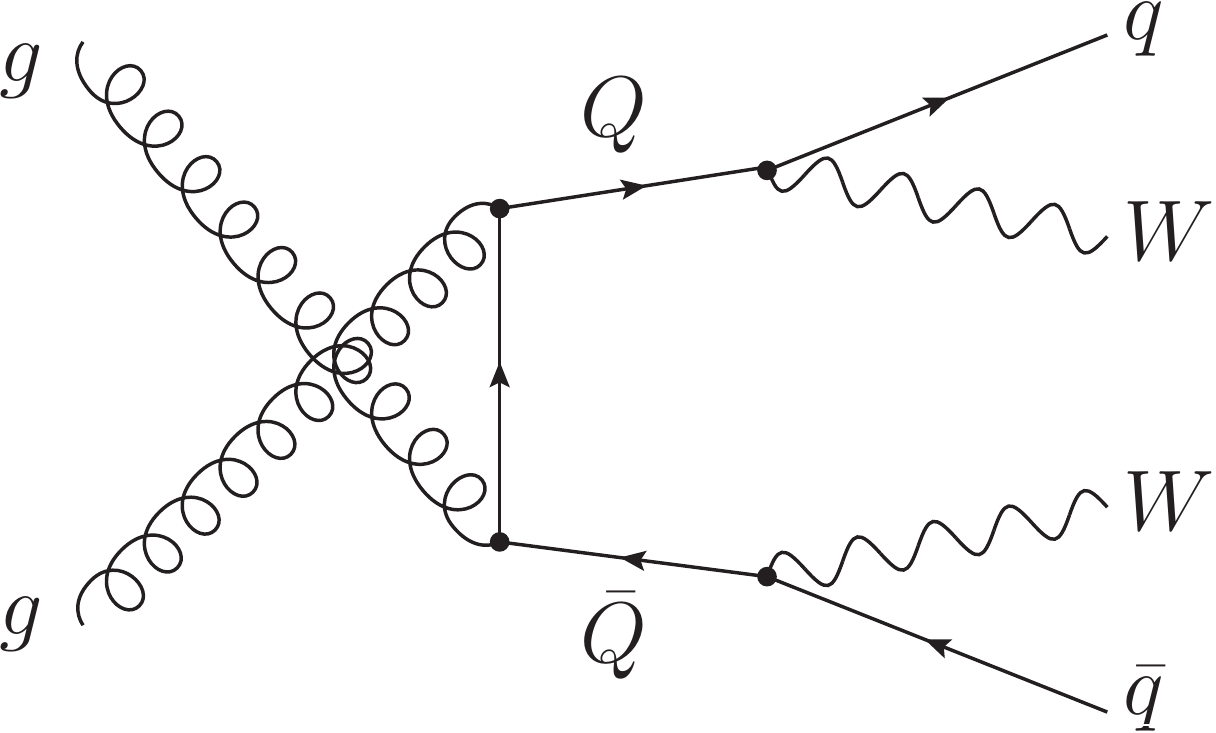}
\hspace{2mm}
\includegraphics[scale=0.28]{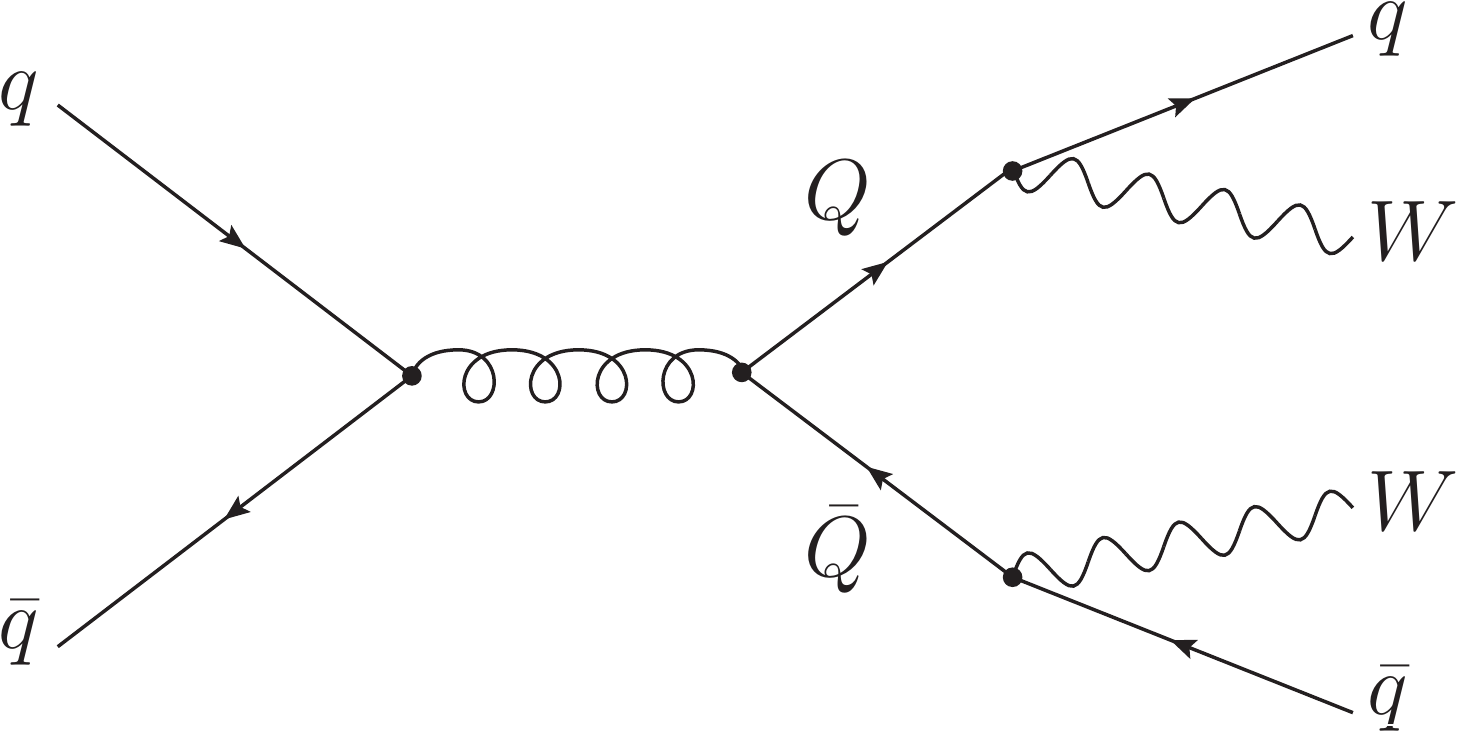}
\caption{Leading-order Feynman diagrams for $Q\bar Q\to WqW\bar q$
  production at the LHC.
\label{fig:feynman}}
\end{center}
\end{figure}

This paper presents a search for new heavy quarks that couple to
light SM quarks using data collected by the ATLAS detector.
The analysis focuses on model-independent pair production of a new
heavy quark and its antiparticle, which then decay through a
charged-current interaction to a final state with a single electron or
muon, missing transverse momentum and light SM quarks, making it complementary to searches for
VLQs that decay to third-generation quarks~\cite{Aad:2015kqa,
  Aad:2014efa, Aad:2015mba, Aad:2015gdg, Chatrchyan:2012vu,
  CMS:2012ab, Chatrchyan:2012af, Chatrchyan:2011ay}.
 The dominant Feynman
diagrams for the signal process are shown in Fig.~\ref{fig:feynman}.
Background rejection is achieved through event topology and kinematic
requirements.  In particular, a kinematic variable motivated by the
splitting scale of a heavy object into daughter
particles~\cite{Butterworth:2002tt,Butterworth:2008iy}
is used to reduce the background when selecting the pair of jets
consistent with a hadronically decaying $W$ boson.  The statistical
interpretation of the data uses the invariant mass distribution of the
$Q$ candidate formed from the hadronically decaying $W$ boson
and a light quark. The results are interpreted both in the context of
VLQs and a chiral fourth-generation quark.

\section {The ATLAS detector}
\label {sec: detector}

The ATLAS detector~\cite{Aad:2008zzm} at the LHC covers nearly the
entire solid angle around the interaction point (IP).\footnote{ATLAS uses a right-handed coordinate system with its
  origin at the nominal IP in the center of the detector and the
  $z$-axis along the beam pipe. The $x$-axis points from the IP to the
  center of the LHC ring, and the $y$-axis points upwards.
  Cylindrical coordinates $(r,\phi)$ are used in the transverse plane,
  $\phi$ being the azimuthal angle around the $z$-axis. The
  pseudorapidity is defined in terms of the polar angle $\theta$ as
  $\eta = -\ln \tan(\theta/2)$. Angular distance is measured in units
  of $\Delta R \equiv \sqrt{(\Delta\eta)^{2} + (\Delta\phi)^{2}}$.}
   It consists of
an inner tracking detector surrounded by a thin superconducting
solenoid, electromagnetic and hadronic calorimeters, and a muon
spectrometer incorporating three large superconducting toroid magnets.

The inner tracking detector system is immersed in a 2\,T axial
magnetic field and provides charged-particle tracking in the range
$|\eta| < 2.5$.  A high-granularity silicon pixel detector covers the
interaction region and typically provides three measurements per
track.  It is followed by a silicon microstrip tracker, which usually
provides four two-dimensional measurement points per track.  These
silicon detectors are complemented by a transition radiation tracker,
which enables radially extended track reconstruction up to $|\eta| =
2.0$.  The transition radiation tracker also provides electron
identification information based on the fraction of hits (typically 30
in total) above a higher threshold for energy deposits corresponding
to transition radiation.

The calorimeter system covers the pseudorapidity range $|\eta| < 4.9$.
Within the region $|\eta|< 3.2$, electromagnetic calorimetry is
provided by barrel and endcap high-granularity lead/liquid-argon (LAr)
 calorimeters, with an additional thin LAr presampler
covering $|\eta| < 1.8$, to correct for energy loss in material
upstream of the calorimeters.  Hadronic calorimetry is provided by the
steel/scintillator-tile calorimeter, segmented into three barrel
structures within $|\eta| < 1.7$, and two copper/LAr hadronic endcap
calorimeters with $|\eta| < 3.2$.  The solid angle coverage is
completed with forward copper/LAr and tungsten/LAr calorimeter modules
optimized for electromagnetic and hadronic measurements respectively.

The muon spectrometer comprises separate trigger and high-precision
tracking chambers measuring the deflection of muons in a magnetic
field generated by superconducting air-core toroids.  The precision
chamber system covers the region $|\eta| < 2.7$ with three layers of
monitored drift tubes, complemented by cathode strip chambers in the
forward region, where the background is highest.  The muon trigger
system covers the range $|\eta| < 2.4$ with resistive plate chambers
in the barrel, and thin-gap chambers in the endcap regions.

A three-level trigger system~\cite{Aad:2012xs} is used to select
events for offline analysis.  The Level-1 trigger is implemented in
hardware and uses a subset of detector information to reduce the event
rate to at most 75 kHz.  This is followed by two software-based
trigger levels that together reduce the event rate to about 400 Hz.

\section{Signal and background samples}
\label{sec:modeling}

The pair-production cross section for a new heavy quark ranges from
12\,pb for a 300$\GeV$ quark to 21\,fb for a 800$\GeV$ quark.  It was
calculated at next-to-next-to-leading-order (NNLO) in QCD, including
resummation of next-to-next-to-leading logarithmic (NNLL) soft gluon
terms, with Top++2.0~\cite{Baernreuther:2012ws, Czakon:2012zr,
  Czakon:2012pz, Czakon:2013goa, Cacciari:2011hy, Czakon:2011xx},
using the MSTW2008 NNLO~\cite{Martin:2009iq, Martin:2009bu} set of
parton distribution functions (PDFs) and $\alphas = 0.117$. The PDF
and $\alphas$ uncertainties were calculated using the PDF4LHC
prescription~\cite{Botje:2011sn} with the MSTW2008 68\% CL NNLO, CT10
NNLO~\cite{Lai:2010vv}, and NNPDF2.3 5f FFN~\cite{Ball:2012cx} PDF
sets.
The uncertainties in the prediction stem from scale variations and the
PDF+$\alphas$ uncertainty and range from approximately 11\% to 12\%
for masses from 300$\GeV$ to 800$\GeV$.

VLQ signal samples were simulated with the tree-level event generator
{\sc CompHEP} v4.5.1~\cite{Boos:2004kh} at the parton level with the
CTEQ6L1 LO PDF set~\cite{Nadolsky:2008zw} and with the QCD scale set
to the mass of the heavy quark, $\MQ$.  The generated events were then
passed into {\sc Pythia} v8.165~\cite{Sjostrand:2007gs,
  Sjostrand:2006za} for hadronization and parton showering.  The VLQ
signal samples were produced for values of $\MQ$ ranging from
300$\GeV$ to 800$\GeV$ in 100$\GeV$ steps.  This range is motivated by
the previous limit at 350$\GeV$ and the expected sensitivity of this
analysis to masses up to approximately 700$\GeV$.  Although the
analysis is targeting the $Q\to Wq$ decay, there is also sensitivity
to the neutral-current decays $Q\to Zq$ and $Q\to Hq$ (e.g., $Q\bar
Q\to WqZ\bar{q}\to\ell\nu qq\bar{q}\bar{q}$) and events were generated
for all six decay combinations.  In addition to the VLQ signal
samples, a set of fourth-generation chiral-quark signal samples were
generated with {\sc Pythia} v8.1 using the MSTW 2008 LO PDF set.  The
kinematics of the chiral-quark signal samples, which only contain
$Q\bar Q\to WqW\bar{q}$, are compatible with the VLQ samples when
requiring BR$(Q\to Wq)=1$.  Therefore, the more generic VLQ samples
are used for the statistical analysis, with the sample corresponding
to $\MQ = 700\GeV$ and BR$(Q\to Wq)=1$ used to represent the signal in
tables and figures, unless noted otherwise.

The background originates mainly from $W$-boson production in association with jets,
$W$+jets, with lesser contributions from top quark pair production ($t\bar{t}$), $Z$+jets,
single-top, diboson, and multijet events.  The $W$+jets and
$Z$+jets samples were produced using {\sc Sherpa} v1.4.1~\cite{Sherpa} with the CT10
PDF set, taking the $c$- and $b$-quarks as massive.  
Samples of $t\bar{t}$ and single-top events were generated with {\sc
  Powheg-Box} 3.0~\cite{Frixione:2007vw,Alioli:2010xd} interfaced to
{\sc Pythia} v6.426~\cite{Sjostrand:2006za} using the Perugia2011C set
of tunable parameters~\cite{Skands:2010ak} for the underlying event
and the CTEQ6L1 PDF set.
Diboson production was modeled using {\sc
  Alpgen} v2.13~\cite{Mangano:2002ea} interfaced to {\sc
  Herwig} v6.520~\cite{Herwig} with the CTEQ6L1 PDF set, except for the
lepton+jets final state ($WV\to \ell\nu qq$ with $V=W,Z$), which used
the {\sc Sherpa} v1.4.1 event generator with the CT10 PDF set.  The
contribution from multijet events originates from the
misidentification of a jet or a photon as an electron, or from the
semileptonic decay of a $b$- or $c$-quark, and the matrix
method~\cite{8TeVFakesConf} is used to determine the kinematic
distributions for the multijet background.

The $W/Z$+jets and multijet backgrounds are normalized to data and a
data-driven correction is applied to the transverse momentum $\pt$
of the boson as described in Sec.~\ref{sec:wjets_model}.  
The $t\bar{t}$ cross section is determined by the Top++ prediction,
 with the top-quark mass taken to be 172.5~\GeV.
The single-top samples are
normalized to the approximate NNLO theoretical cross
sections~\cite{Kidonakis:2011wy, Kidonakis:2010ux, Kidonakis:2010tc}
calculated using the MSTW 2008 NNLO PDF set.  The diboson background
processes are normalized to NLO theoretical cross
sections~\cite{Campbell:1999ah}.

All simulated samples include
multiple $pp$ interactions per bunch crossing and simulated events are weighted such that
the distribution of the average number of interactions per bunch
crossing agrees with data. The generated samples are processed through
a simulation \cite{Aad:2010ah} of the detector geometry and response
using {\sc Geant4}~\cite{Agostinelli:2002hh}, then reconstructed using
the same software as used for data.  Simulated events are corrected so
that the object identification efficiencies, energy scales,
and energy resolutions match those determined in data control samples.

\section{Event selection}
\label {sec:data}

The data analyzed in this search were collected with the ATLAS
detector in 2012 and correspond to an integrated luminosity of
20.3\,\ifb.  Data quality requirements are applied to remove events
with incomplete, corrupted, or otherwise compromised subdetector information.
Events are required to pass a single-electron or single-muon trigger.
The $\pt$ thresholds are $24\GeV$ and $60\GeV$ for the electron
triggers and $24\GeV$ and $36\GeV$ for the muon triggers. The
lower-threshold triggers include isolation requirements on the
candidate leptons, resulting in inefficiencies at higher $\pt$ that
are recovered by the higher-threshold triggers without an isolation
requirement.

\subsection{Preselection}

The basic object selection is called preselection and requires exactly
one charged lepton (electron or muon), at least four jets, and large
missing transverse momentum (${E}_\text{T}^\text{miss}$), as described
below.  The criteria are similar to those used in recent ATLAS
top-quark studies~\cite{Aad:2015pga}, except that this analysis
requires that there are no jets identified as originating from a
$b$-quark.  The expected and observed numbers of events after
preselection are shown in Table~\ref{presel}.  There is negligible
sensitivity to heavy quark production because the signal expectations
for all masses are much smaller than the uncertainty on the
background, dominated at this stage by systematic uncertainties.

\subsubsection*{ Charged-lepton requirements}
Electron candidates are required to have $\pt > 25\GeV$ and either
$|\eta| < 1.37$ or $1.52 < |\eta| < 2.47$ to exclude the transition
between the barrel and endcap calorimeters.
Muon candidates are required to have $\pt > 25\GeV$ and $|\eta| < 2.5$.
Nonprompt leptons and nonleptonic particles may be reconstructed as
leptons and satisfy the selection criteria, giving rise to nonprompt
and fake lepton backgrounds. In the case of electrons, these include
contributions from semileptonic decays of $b$- and $c$-quarks, photon
conversions, and jets with a large electromagnetic energy fraction.
Nonprompt or fake muons can originate from semileptonic decays of $b$-
and $c$-quarks, from charged-hadron decays in the tracking volume or
in hadronic showers, or from punchthrough particles emerging from
high-energy hadronic showers~\cite{8TeVFakesConf}. The nonprompt and
fake lepton backgrounds are reduced by requiring the lepton candidates
to be isolated from other energy deposits or high-\pt{} tracks.  The
tracks used in the isolation calculation are required to originate
from the primary interaction vertex and have $\pt > 1\GeV$.  For
electrons, an $\eta$-dependent limit is placed on the amount of energy
measured in the calorimeter within a $\Delta R=0.2$ cone around the
candidate which is neither from the electron candidate itself nor from
additional $pp$ interactions.  A similar requirement is placed on the
scalar sum of the $\pt$ of tracks within a $\Delta R=0.3$ cone around
the track of the electron candidate.  Each requirement has an average
efficiency of 90\% for electrons from $Z\to ee$.  Muons are required
to have a \pt-dependent track isolation~\cite{Rehermann:2010vq},
requiring the scalar sum of the $\pt$ from tracks with $\Delta
R<10\GeV/\pt^{\mu}$ to be less than $0.05\cdot\pt^\mu$, where
$\pt^\mu$ is the $\pt$ of the candidate muon track.  This requirement
has an efficiency of approximately 95\% for muons from $W\to\mu\nu$.

In this analysis, $\tau$ leptons are not explicitly reconstructed.
Because of the high $\pt$ threshold, only a small fraction of $\tau$
leptons decaying leptonically are reconstructed as electrons or muons,
while the majority of $\tau$ leptons decaying hadronically are
reconstructed as jets.

\subsubsection*{ Jet requirements }
Events must contain at least four jets with $\pt>25\GeV$ and
$|\eta|<2.5$ reconstructed using the anti-$k_{t}$
algorithm~\cite{Cacciari:2008gp} with a radius parameter $R=0.4$.  The
jets are constructed from calibrated topological clusters built from
energy deposits in the calorimeters, and they are calibrated to the
hadronic scale~\cite{Aad:2014bia}.  Prior to jet finding, a local
cluster calibration scheme~\cite{Barillari:2009zza} is applied to
correct the topological cluster energies for the effects of calorimeter 
noncompensation, dead material and out-of-cluster leakage.  
The jet energy scale was determined using information from
test-beam data, LHC collision data, and simulation~\cite{Aad:2011he, Aad:2014bia}.  
All jets are required to have
at least two tracks that originate from the primary interaction
vertex. In order to suppress jets that do not originate from the
primary vertex, jets with $\pt < 50\GeV$ and $|\eta| < 2.4$ are
required to have a jet vertex fraction (JVF) above 0.5, where JVF is
the ratio of the summed scalar $\pt$ of tracks originating from the
primary vertex to that of all tracks associated with the jet.  An
overlap removal procedure~\cite{Aad:2015pga} is applied to remove jets
that were already identified as electrons.

To identify jets originating from the hadronization of a $b$-quark
(`$b$-tagging'), a continuous discriminant is produced by an algorithm
using multivariate techniques~\cite{bcltag} to combine information
from the impact parameter of displaced tracks and topological
properties of secondary and tertiary vertices reconstructed within the
jet.  The efficiency for a jet containing a $b$-hadron to be
$b$-tagged is 70\%, while the light-jet ($c$-jet) efficiency is less
than 1\% (20\%) as determined in simulated $t\bar{t}$ events, where
light jets are jets initiated by a $u$-, $d$-, $s$-quark, or gluon.
If any jets are $b$-tagged, the event is rejected.

\subsubsection*{Missing transverse momentum requirements} 
The $\vec{E}_\text{T}^\text{miss}$ is constructed from the vector sum
of calibrated energy deposits in the calorimeter and reconstructed
muons~\cite{met}.  Events that do not contain a leptonically decaying
$W$ boson are suppressed by requiring $\met > 30\GeV$ and $(\met +
\mt) > 60\GeV$, where $\mt$ is the transverse mass of the $W$ boson
defined as $\mt=\sqrt { 2 \pt^{\ell} \met (1-\cos \Delta\phi) }$,
where $\Delta\phi$ is the azimuthal angle between the charged-lepton
transverse momentum vector and $\vec{E}_\text{T}^\text{miss}$.

\begin{table}[htbp]
  \begin{center}
    \caption{\label{presel} Observed and expected event yields after
      preselection.  The quoted uncertainties include both the
      statistical and systematic contributions, with the latter being
      dominant. The sources of systematic uncertainty are discussed in
      Sec.~\ref{sec:systematics}. }

\renewcommand{\arraystretch}{1.4}
\begin{tabular}{ l*{2}{r@{ }l} }

\hline
\hline
  & \multicolumn{2}{c}{ Electron} & \multicolumn{2}{c}{ Muon}\\
\hline

$W$+jets & $110000$ & $^{+15000}_{-21000}$ & $145000$ & $^{+20000}_{-28000}$ \\

$Z$+jets & $28000$  & $^{+14000}_{-15000}$ & $15200$ & $^{+7700}_{-7800}$\\

$t\bar{t}$ & $19200$  & $^{+2800}_{-2900}$ & $23700$ & $^{+3400}_{-3500}$ \\

Diboson & $3900$  & $\pm 1900$ & $4400$ & $\pm 2200$ \\

Single top & $3750$  & $^{+540}_{-560}$ & $4500$  & $^{+630}_{-710}$ \\

Multijet & $22000$  & $ \pm 8800$ & $13300$ & $ \pm 5300$ \\

\hline
Total background & $183000 $  & $ ^{+27000}_{-33000}$ & $206000$  & $ ^{+27000}_{-34000}$ \\
\hline

Signal ($\MQ=700\GeV$) & $79$ & $ \pm 10$ & $79$  & $\pm 10$ \\ 

\hline

Data & \multicolumn{2}{c}{$182075$} & \multicolumn{2}{c}{$208641$} \\
\hline
\hline
\end{tabular}

  \end{center}
\end{table}

\subsection{Final selection}
\label{sec:final_sel}
With the final-state objects identified, additional kinematic
requirements are applied to exploit the distinct features of the
signal, assumed to be heavier than the previously excluded mass of
$350\GeV$.  The $W$ boson and light quark originating from $Q\to Wq$
would be very energetic and have a large angular separation due to the
large mass of the $Q$ quark.  On the other hand, the decay products of
the $W$ boson would tend to have a small separation due to the $W$
boson's boost.  By selecting only events that are consistent with
these properties, the $W$+jets background yield is reduced by orders
of magnitude.  To facilitate the discussion of the kinematic
selection, the following objects are defined:
\begin{itemize}
  \item $\Whad$ is the candidate for a $W$ boson in the decay $Q\to  
  Wq\to qqq$;
  \item $\Wlep$ is the candidate for a $W$ boson in the decay $Q\to
    Wq\to\ell\nu q$;
  \item $q$-jet is a candidate for the jet originating from the $q$ in
    $Q\to Wq$ (i.e., from the decay of the heavy quark, not the
    hadronically decaying $W$ boson).  There are two $q$-jets per event,
    so $q_1$ ($q_2$) is used to denote the one with higher (lower)
    \pt.
\end{itemize}

The $\Whad$ candidate is defined as a dijet system with $\pt >
200\GeV$, angular separation $\Delta R < 1.0$, and an invariant mass
in the range 65--100$\GeV$.  All possible jet combinations are
considered and, if multiple pairs satisfy the above requirements, the
pair with the mass closest to that of the $W$ boson~\cite{PDG} is
chosen.  If no $\Whad$ candidate is found, the event is removed.  This
requirement results in 94\% background rejection while keeping 53\% of
the signal if $m_Q = 700\GeV$.  The mass distribution of the $\Whad$
candidates prior to the mass requirement is shown in
Fig.~\ref{fig:dijetinvariantmass}.

\begin{figure}[htbp]
\begin{center}
\begin{tabular}{cc}
\includegraphics[width=0.48\textwidth]{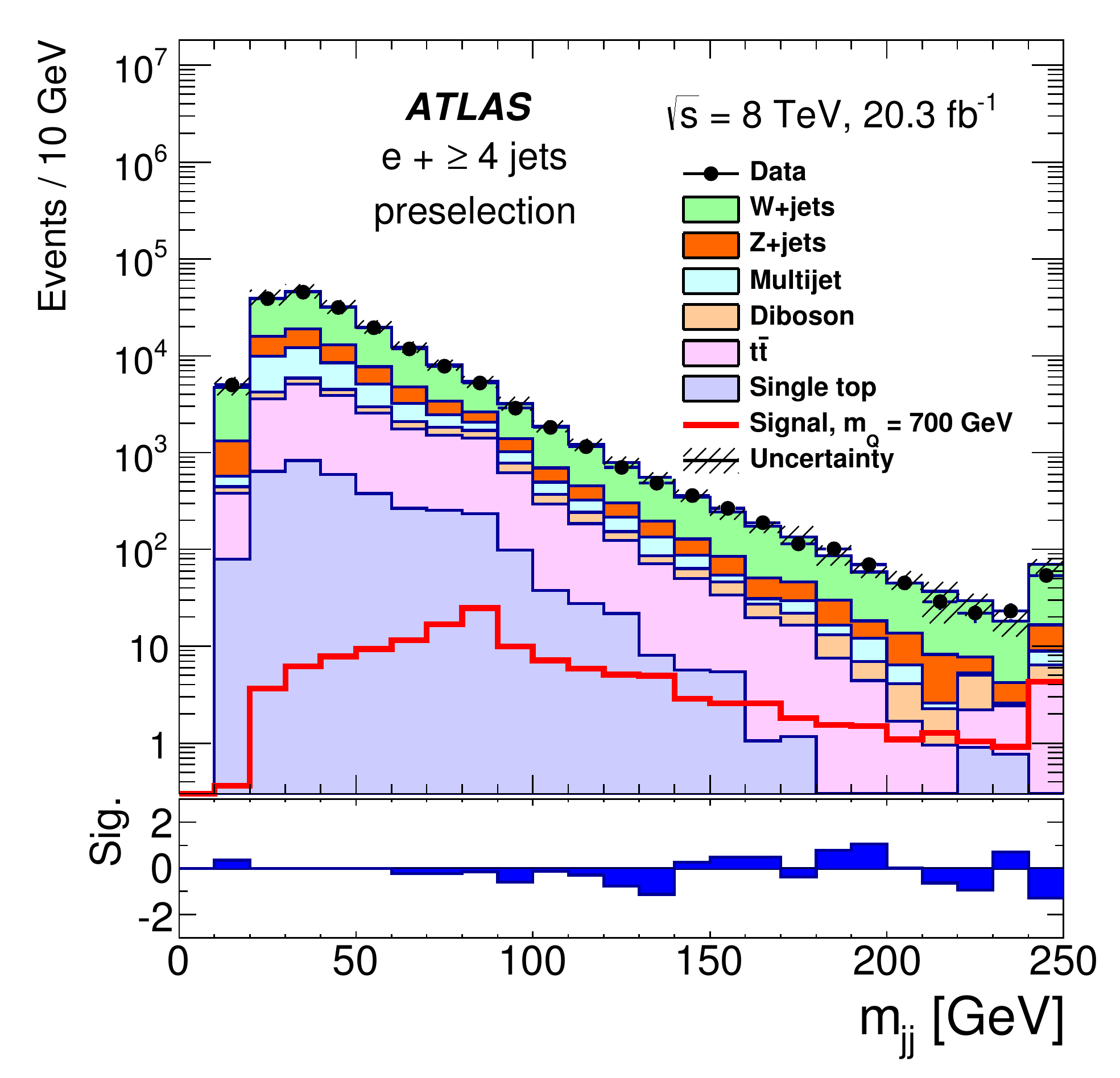}
\includegraphics[width=0.48\textwidth]{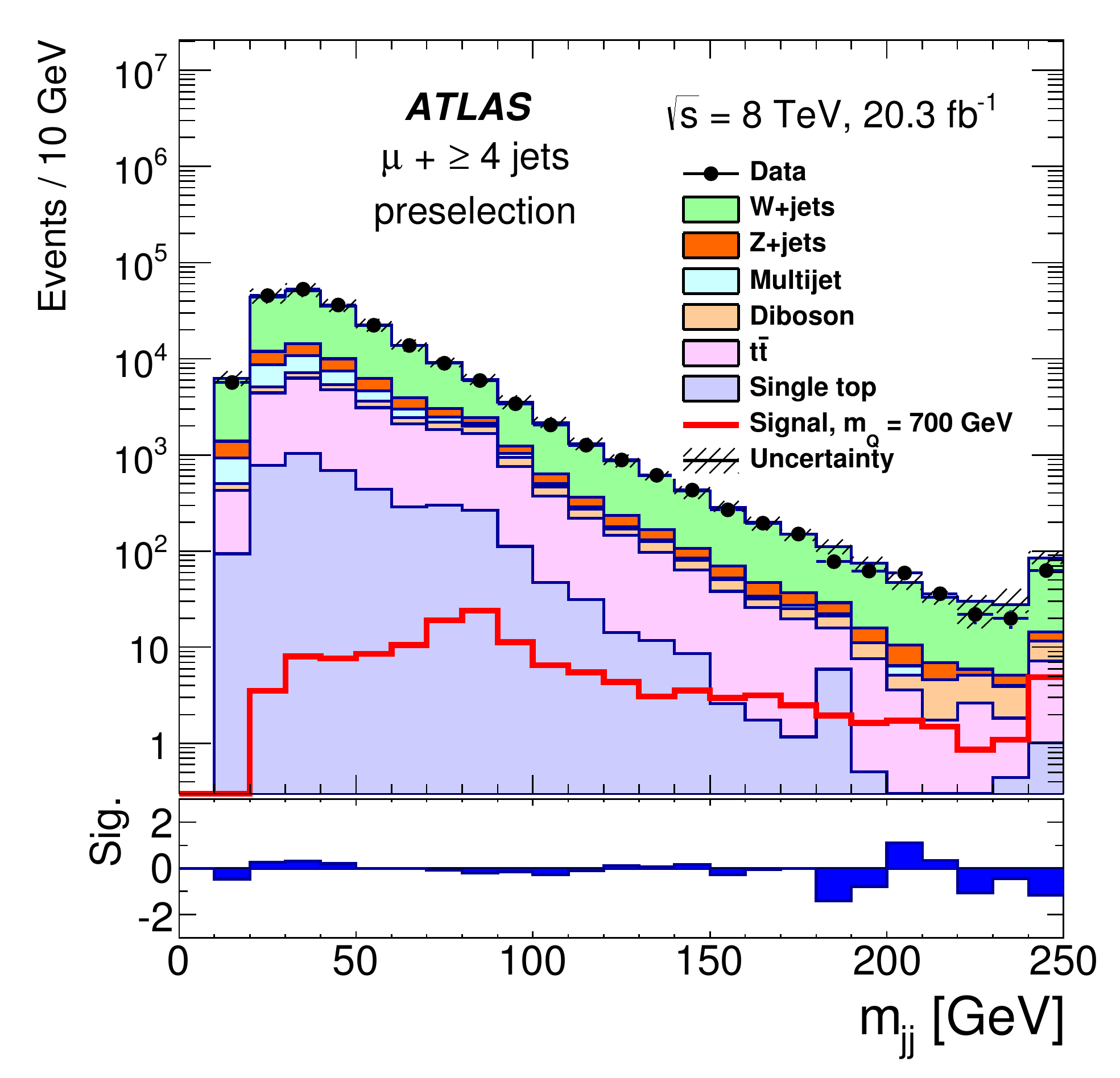}
\end{tabular}
\caption{Invariant mass distributions for dijet systems with $\pt >
  200\GeV$, angular separation $\Delta R < 1.0$, and an invariant mass
  closest to the $W$ boson mass, for the (left) electron and (right) muon
  channels after the preselection requirements.  $\Whad$ candidates
  correspond to the mass range 65--100$\GeV$. The highest bin includes all
  events with $m_{jj}>240\GeV$.  
  The shaded band shows the total
  uncertainty on the background prediction.  The lower panel shows the
  significance of the difference between data and expectation in
  units of normal standard deviations as explained in
  Ref.~\cite{statplots:1}.
\label{fig:dijetinvariantmass}}
\end{center}
\end{figure}

The $\Wlep$ candidate is reconstructed using the lepton and $\vec{E}_\text{T}^\text{miss}$,
which is taken to be the neutrino $\vec{p}_\text{T}$.  The longitudinal momentum of the
neutrino is determined up to a two-fold ambiguity by requiring the
invariant mass of the lepton--neutrino system to equal the mass of the
$W$ boson. When no real solution exists, the neutrino pseudorapidity
is set equal to that of the lepton because the decay products of the
$W$ boson tend to be nearly collinear for the kinematic regime of
interest. In simulated samples, the rate of events with no real solution is approximately 35\%. 
 Signal events are expected to have energetic $W$ bosons, so
the $\Wlep$ candidate is required to have $\pt > 125\GeV$.

The $W$ candidates ($\Whad$ and $\Wlep$) are then each paired with a
different one of the remaining jets to create the two heavy quark
candidates, $Q1$ and $Q2$.  This step involves testing all possible pairings of
$q$-jet candidates with the $\Whad$ and $\Wlep$ candidates.  In
addition, the $\Wlep$ candidate may have two real solutions for the
longitudinal momentum of the neutrino.  Among the possible
combinations of neutrino momentum solutions and $Wq$ pairings, the one
yielding the smallest absolute difference between the two
reconstructed heavy quark masses is chosen.  In simulated samples, the
rate of correct $Wq$ pairing is approximately 40\% (48\%) for a signal
of mass 400~\GeV~(800~\GeV).  Once the heavy quark candidates are
determined, the $q$-jets are required to have $\pt(q_1) > 160\GeV$ and
$\pt(q_2) > 120\GeV$ and the difference between the reconstructed
heavy quark masses must be less than 120$\GeV$.

With candidate objects identified for two heavy quarks and their decay
products, the following additional kinematic criteria are applied.
Each event must have $\HT > 1100\GeV$, where $\HT$ is the scalar sum
of the transverse momenta of the lepton, neutrino, $\Whad$, and the
two $q$-jets.  The angular separation between the lepton and neutrino
candidates must satisfy $\Delta R(\ell, \nu) < 1.4$.  The heavy quarks
would tend to be central and back-to-back, so the angular separation
between the two reconstructed heavy quarks is required to satisfy
$2.0 <  \Delta R(Q1, Q2) <  4.2$.

To further capitalize on the presence of a hadronically
decaying $W$ boson with high $\pt$ in the signal, the analysis makes use of a
splitting variable~\cite{Butterworth:2008iy} defined as:
\[   y_{12} = \frac{\min(p_{\text{T}j1},p_{\text{T}j2})^2 \times \Delta R(j_1,j_2)^2}{m_{j1j2}^2};
\]
\noindent where $j_1$ and $j_2$ are the two jets from the $\Whad$
candidate and $m_{j1j2}$ is the mass of the $\Whad$ candidate.  The
two jets from a hadronically decaying $W$ boson tend to have roughly
equal energies, while dijets from QCD processes are likely be
asymmetric in energy.  Furthermore, jets from $W\to qq$ tend to have a
larger opening angle due to the mass of the $W$ boson.  Dividing by
the dijet mass provides discrimination for cases in which a QCD dijet
system happens to have large values for
$\min(p_{\text{T}j1},p_{\text{T}j2})^2$ and $\Delta R(j_1,j_2)^2$, as
such systems are likely to have a very large mass. The background is
reduced by a factor of 3.2 by requiring $y_{12} > 0.25$.  This
requirement has a signal efficiency of approximately 50\%, although
the precise value depends on the mass of the heavy quark.  The
modeling of the splitting variable with a large number of events can
be seen in the top histograms of Fig.~\ref{fig:splitting}, which
depict the $y_{12}$ distribution after preselection and requiring a
$\Whad$ candidate.  The bottom of Fig.~\ref{fig:splitting} contains
the distributions of $y_{12}$ immediately before the $y_{12}>0.25$
requirement is applied.

\begin{figure}[htbp]
\begin{center}
\begin{tabular}{cc}
\includegraphics[width=0.48\textwidth]{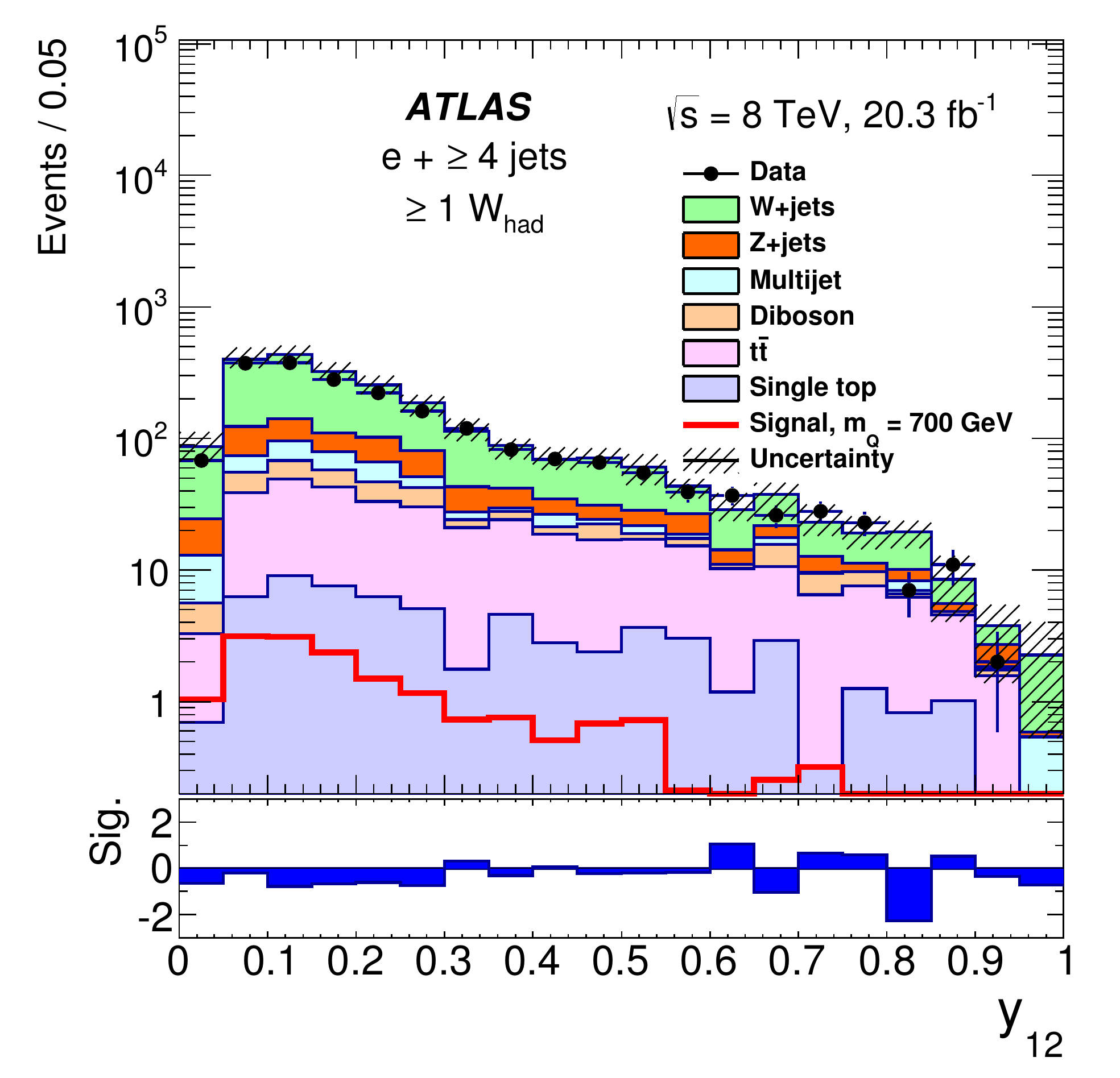}
\includegraphics[width=0.48\textwidth]{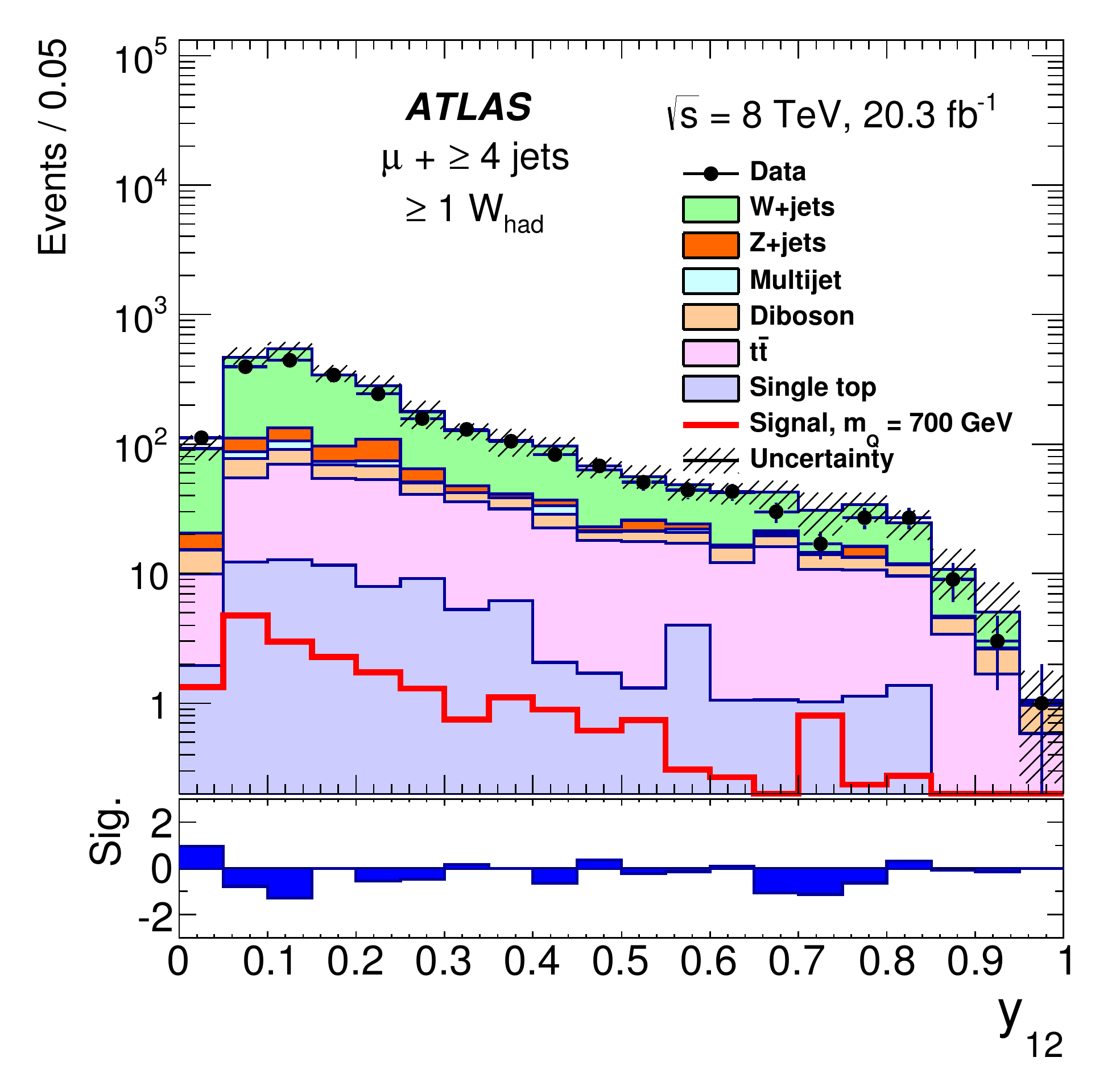}\\
\includegraphics[width=0.48\textwidth]{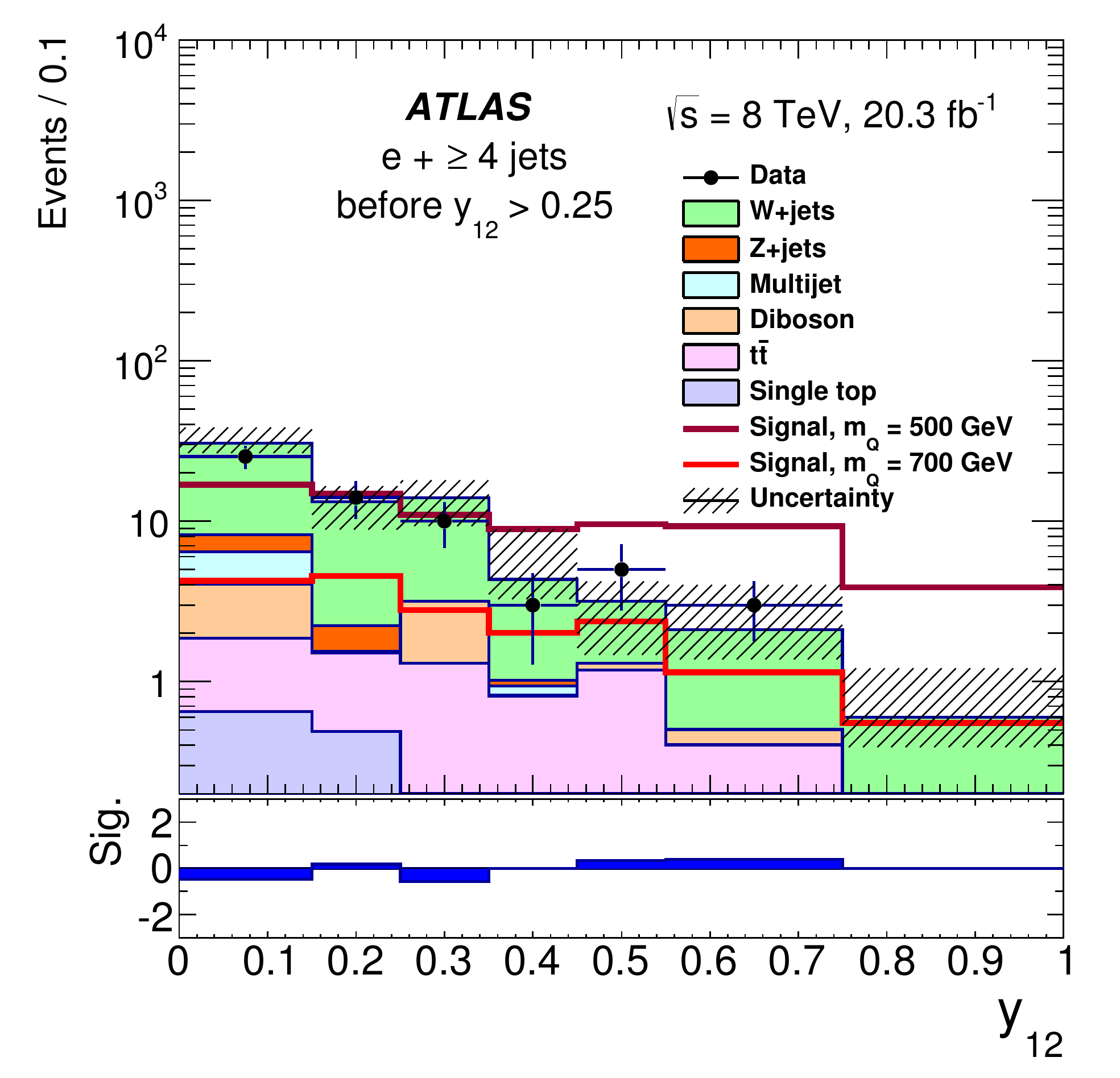}
\includegraphics[width=0.48\textwidth]{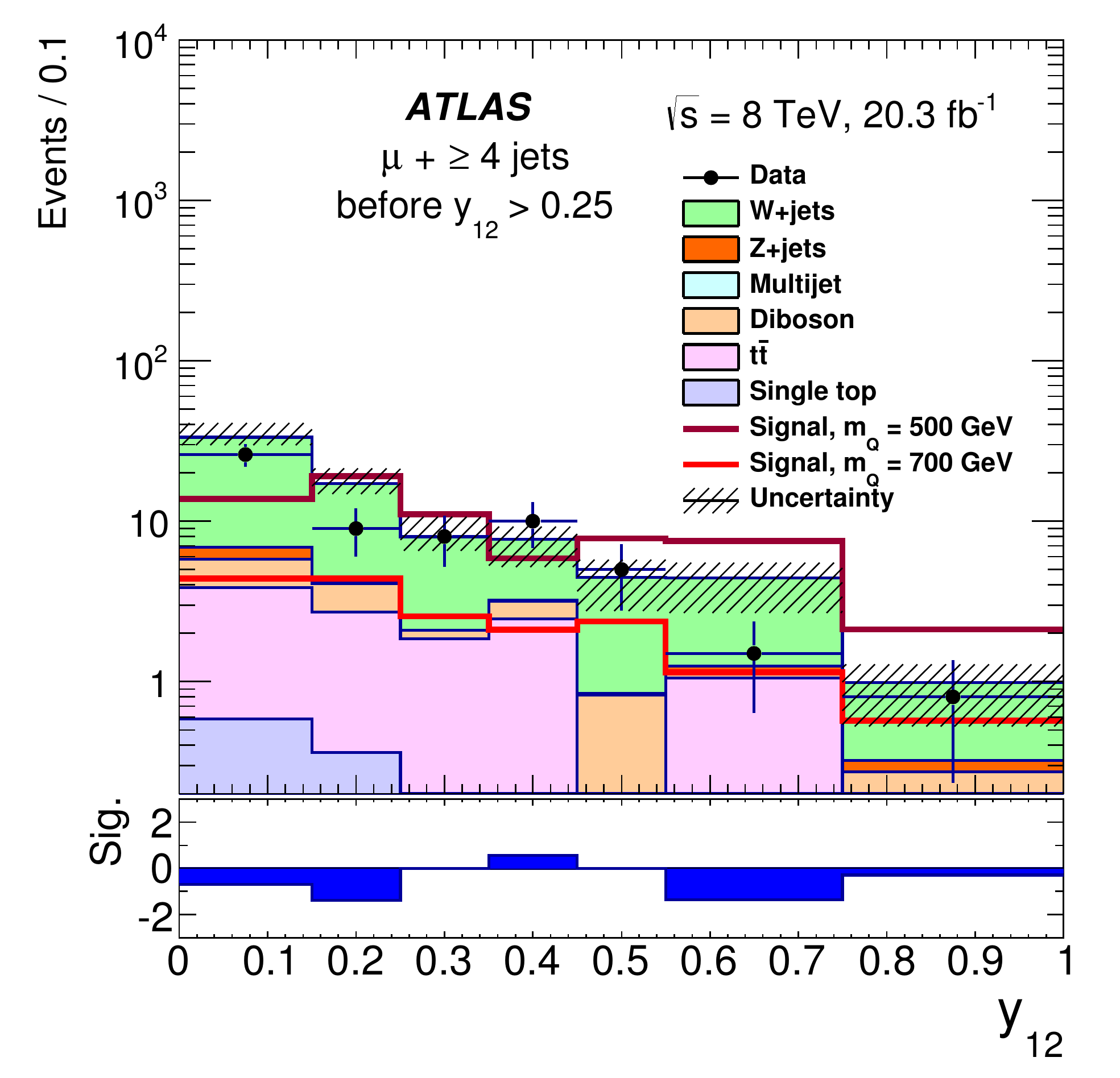}
\end{tabular}
\caption{The splitting variable $y_{12}$ between the decay products of
  the hadronic $W$ boson (top) after preselection and requiring a
  hadronic $W$ boson candidate and (bottom) immediately before
  applying the requirement $y_{12}>0.25$, for the (left) electron and
  (right) muon channels.  The shaded band shows the total uncertainty
  on the background prediction.  The lower panel shows the
  significance of the difference between data and expectation in units
  of normal standard deviations.
\label{fig:splitting}}
\end{center}
\end{figure}

The final selection criteria are motivated by the fact that the decay
products from the $Q$ quark are well separated.  The final
requirements are $\Delta R(\Whad, q_{1}) > 1.0$, $\Delta R(\Whad,
q_{2}) > 1.0$, $\Delta R(\Wlep, q_{1}) > 1.0$, and $\Delta R(\Wlep,
q_{2}) > 1.0$.  Table~\ref{tab:totalYields} presents a summary of the
expected and observed numbers of events after the final selection, for
which the signal (background) efficiency compared to preselection is
approximately 8\% (0.004\%). The small contributions from $t\bar{t}$,
$Z$+jets, dibosons, single-top, and multijet events are combined into
a single background source referred to as `non-$W$+jets'.
Uncertainties on the yields include the uncertainty due to the size of
the signal and background samples and the cumulative effect of the
systematic uncertainty described in Sec.~\ref{sec:systematics}.

The final discriminant variable used in this search is \Mreco, the
reconstructed heavy quark mass built from the $\Whad$ candidate and
the paired $q$-jet candidate.

\renewcommand{\arraystretch}{1.3}
\begin{table}[h!]\centering

\caption{Expected yields for the backgrounds and the VLQ signal with
  $m_Q = 700\GeV$, along with the observed number of data events,
  after applying all selection criteria.  The uncertainties on the
  predicted yields correspond to the statistical uncertainty due to
  finite sample size and the systematic uncertainty, respectively. }
\begin{tabular}{ l*{2}{r@{ $\pm$}l@{ }l}}
  \hline\hline
  & \multicolumn{3}{c}{Electron} 
  & \multicolumn{3}{c}{Muon} \\
  \hline

 $W$+jets 
 & $5.6$  & $1.5$ & $^{+1.5}_{-1.2}$
 & $6.0$  & $1.0$ & $^{+2.2}_{-1.6}$\\

  Non-$W$+jets 
 & $1.2$  & $0.5$ & $^{+1.0}_{-0.4}$
 & $1.2$  & $0.4$ & $^{+0.8}_{-1.0}$\\
  \hline

  Total background
 & $6.8 $  & $ 1.6$ & $ ^{+2.4}_{-1.5}$
 & $7.2 $  & $ 1.1$ & $ ^{+2.5}_{-2.3}$\\
  \hline

  Signal ($\MQ=700\GeV$) 
 & $7.0$  & $0.6$ & $^{+1.1}_{-1.3}$
 & $6.9$  & $0.6$ & $^{+1.0}_{-1.0}$\\
  \hline

  Data & \multicolumn{3}{c}{9} & \multicolumn{3}{c}{11} \\

  \hline\hline
\end{tabular}

\label{tab:totalYields} 
\end{table}

\section{Background modeling}

\subsection{Correction to $W$+jets modeling}
\label{sec:wjets_model}

It is observed after applying the preselection criteria that the
simulated $W$+jets sample does not accurately model the $\pt$ spectrum
of the leptonically decaying $W$ boson candidate.  This mismodeling
leads to an overestimation of the $W$+jets yields in the high-momentum
tails of the $\met$, lepton $\pt$, jet $\pt$, and $H_{\rm T}$
distributions.
The dominant background for this analysis is $W$+jets events in which
the transverse momentum of the $W$ boson, $\pt(W)$, is
high. Therefore, it is important to have accurate predictions for both
the overall normalization and the $\pt(W)$ distribution.  This section
describes the procedure used to derive the $W$+jets and multijet
normalizations and a reweighting function to correct the vector-boson
$\pt$ in the $W$+jets and $Z$+jets\footnote{The correction is
  primarily motivated by the dominant $W$+jets background, but is also
  applied to the $Z$+jets background.}  simulated samples.
All steps in the procedure rely on fits to the data using the $\pt$
distribution of the $W_\text{lep}$ candidate, $\pt^\mathrm{reco}(\Wlep)$,
after applying the preselection requirements.

First, the normalizations for the $W$+jets templates and the multijet
templates are fit to the data, with all other background processes
fixed at their expectations.  The difference between the observed and
predicted number of $W$+jets events is assumed to be due to the cross
section differing from its predicted value, so the electron and muon
channels are fit simultaneously to determine a single $W$+jets scale
factor of 0.82.
A correction for $\pt^\text{truth}(V)$ is then derived that minimizes
the $\chi^2$ between data and simulation for the
$\pt^\text{reco}(\Wlep)$ distribution, where $\pt^\text{truth}(V)$ is
the $\pt$ of the generated vector boson in the $W$+jets or $Z$+jets
sample just before its leptonic decay.
The reweighting is approximately unity for low $\pt^\text{truth}(V)$,
decreasing to 0.86 for $\pt^\text{truth}(V)=200\GeV$ and 0.65 for
$\pt^\text{truth}(V)=500\GeV$.  Finally, the normalizations for the
multijet samples and the {\em corrected} $W$+jets samples are fit to
the data, with all other background processes fixed.  The fit is done
simultaneously in the electron and muon channels.
Figure~\ref{wptrw:fig:step3} shows the $\pt^\mathrm{reco}(\Wlep)$
distribution in the electron and muon channels after applying the
$\pt^\text{truth}(V)$ correction and scale factors.  The corresponding
distributions for $\HT$ and $\met$ are shown in
Figs.~\ref{wptrw:fig:HTstep3} and~\ref{wptrw:fig:METstep3}.  The
uncertainties on the normalizations and $\pt^\text{truth}(V)$
reweighting are described in Sec.~\ref{sec:systematics}.

\begin{figure}[!h]
\begin{center}
\includegraphics*[width=0.47\textwidth] {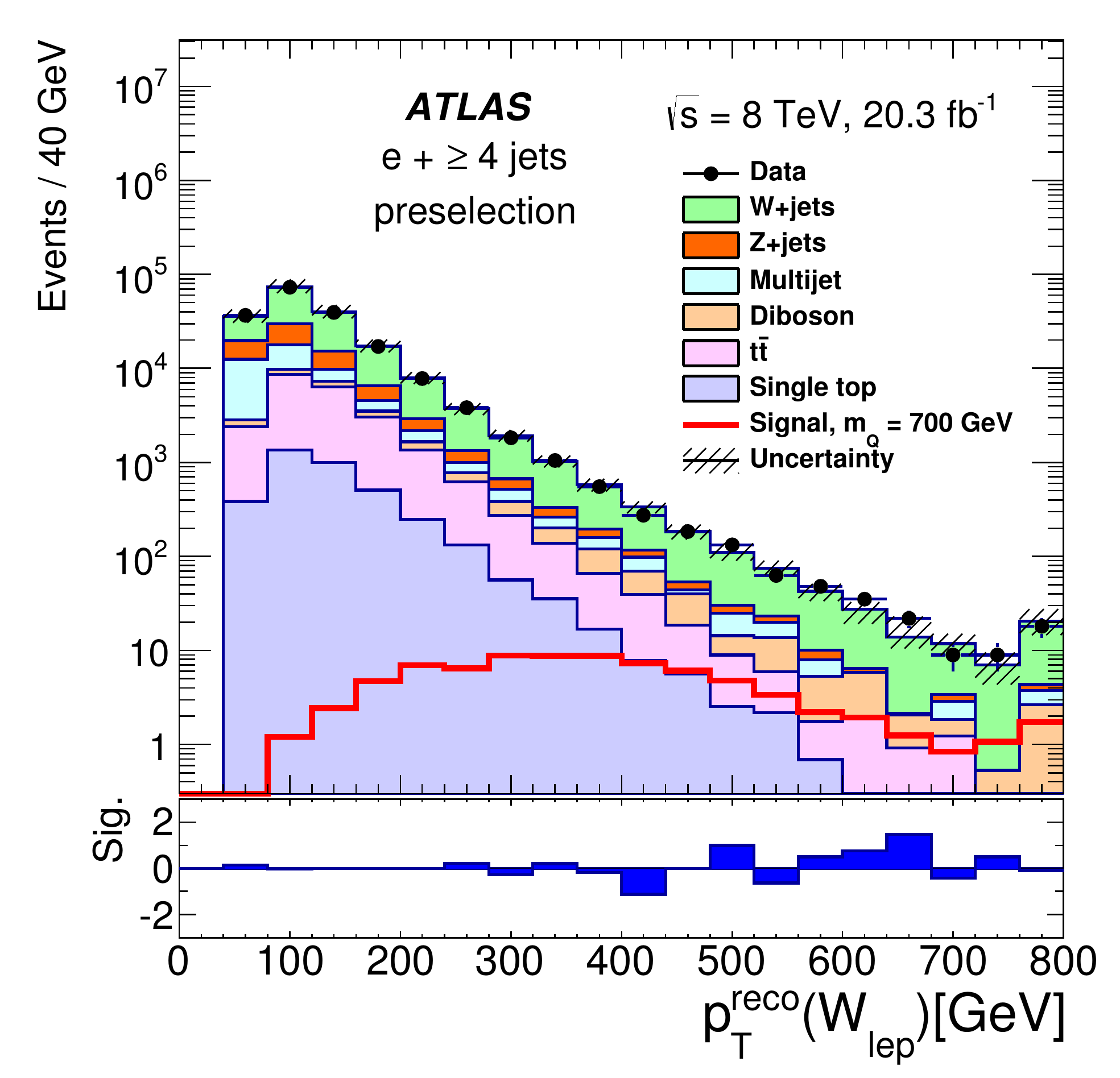}
\includegraphics*[width=0.47\textwidth] {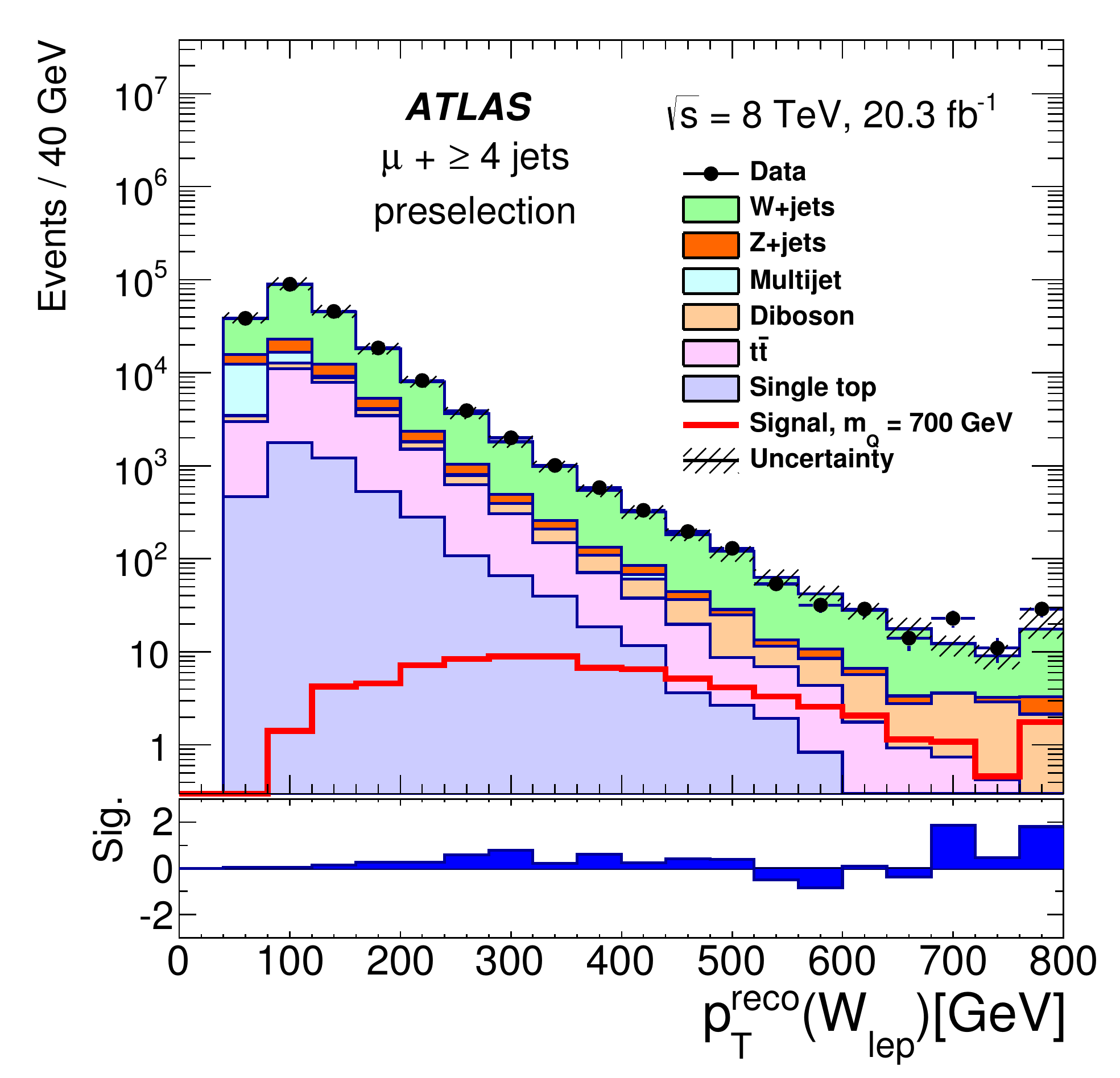}
\caption{The transverse momentum distributions for the leptonic $W$ boson candidate
  after the $\pt^\text{truth}(V)$ correction and final normalization
  for the (left) electron and (right) muon channels.  All overflows
  are included in the rightmost bin.  The shaded band shows the total
  uncertainty on the background prediction.  The lower panel shows the
  significance of the difference between data and expectation in
  units of normal standard deviations.
\label{wptrw:fig:step3}
}
\end{center}
\end{figure}

\begin{figure}[!h]
\begin{center}
\includegraphics*[width=0.47\textwidth] {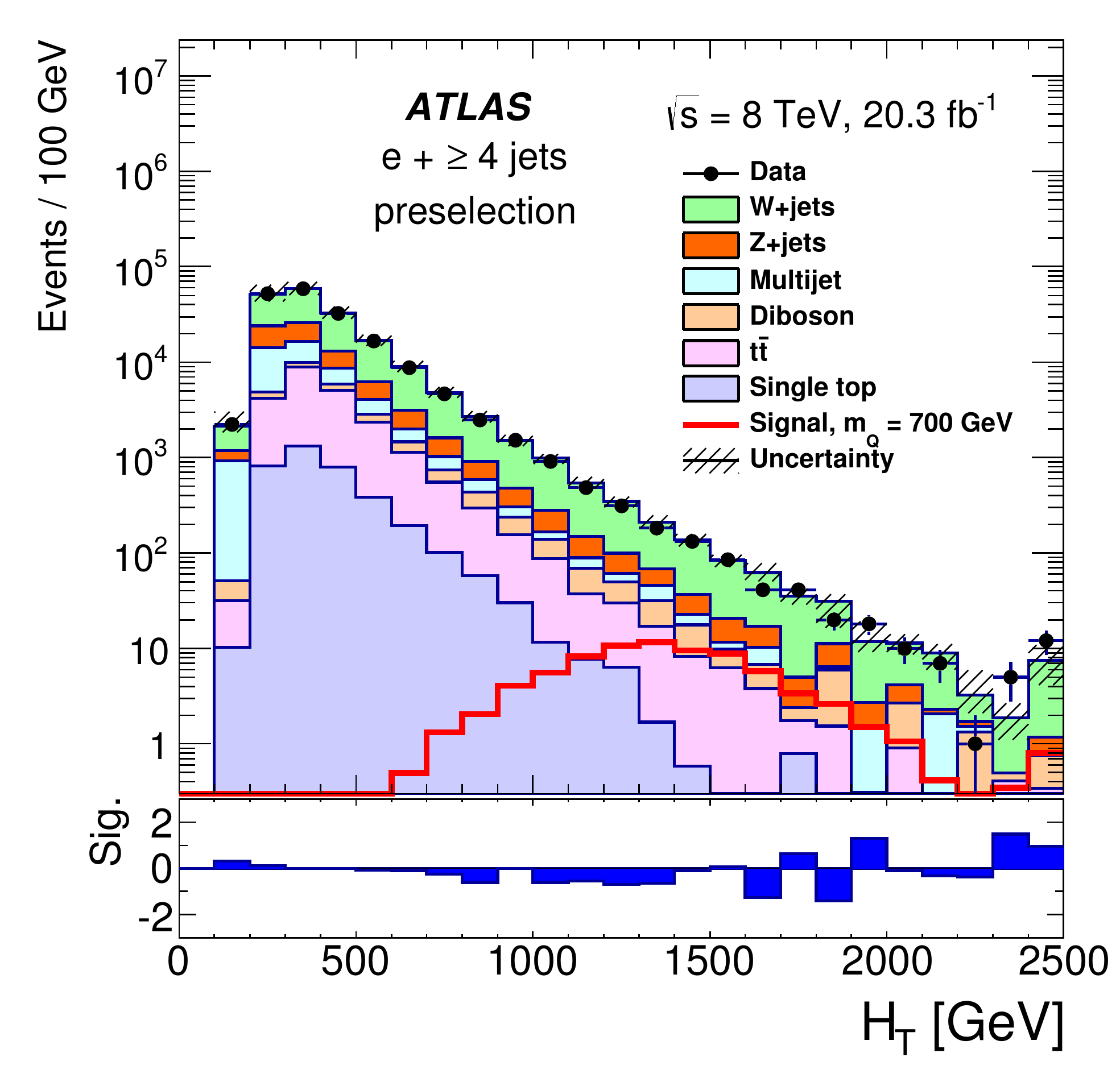}
\includegraphics*[width=0.47\textwidth] {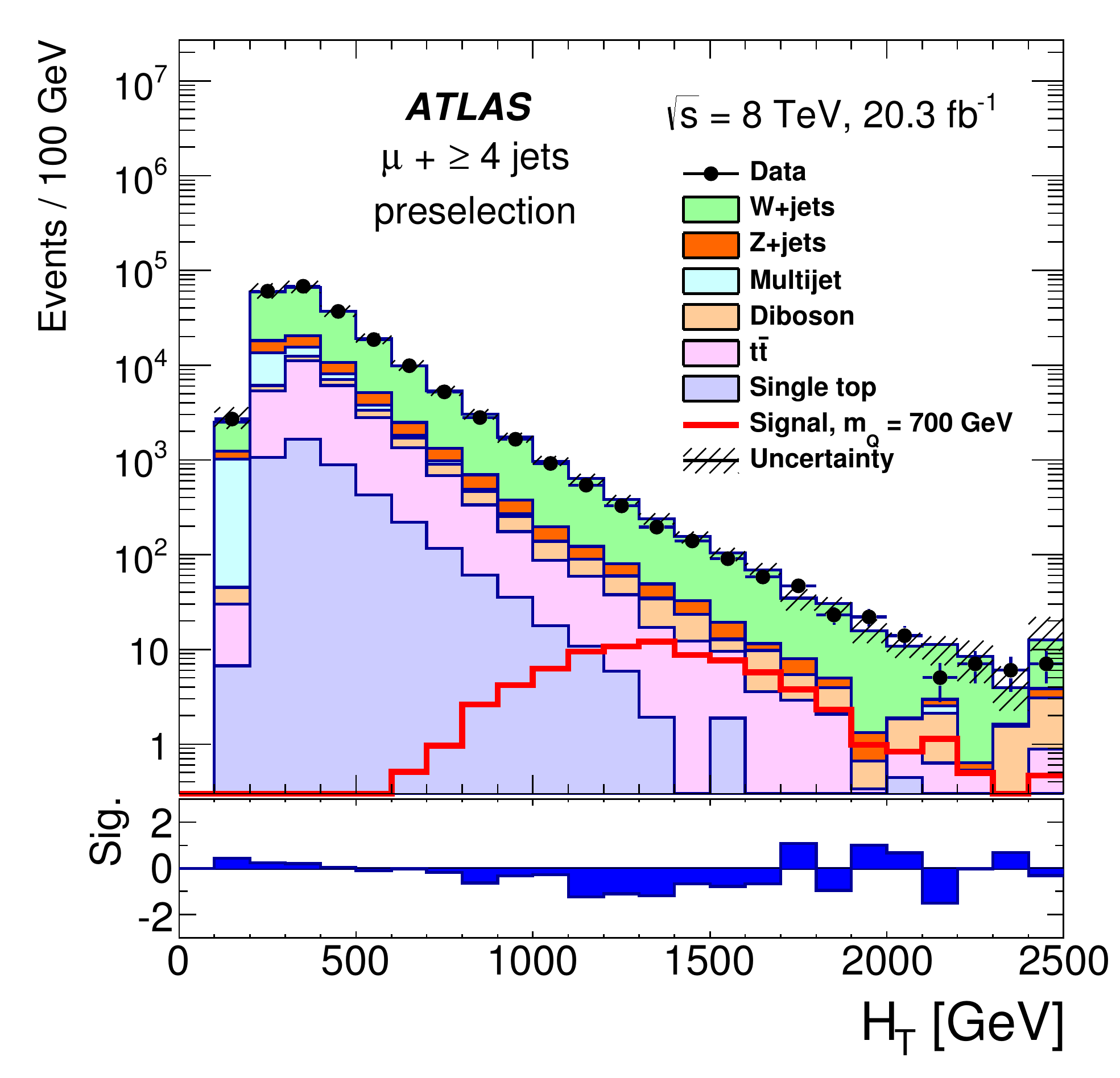}
\caption{Distributions of the scalar sum of the transverse momenta
of the lepton, neutrino, $W_{had}$, and the two $q$-jets ($H_{\rm T}$) after the $\pt^\text{truth}(V)$
  correction and final normalization for the (left) electron and
  (right) muon channels.  All overflows are included in the rightmost
  bin.  The shaded band shows the total uncertainty on the background
  prediction.  The lower panel shows the significance of the
  difference between data and expectation in units of normal
  standard deviations.
\label{wptrw:fig:HTstep3}
}
\end{center}
\end{figure}

\begin{figure}[!h]
\begin{center}
\includegraphics*[width=0.47\textwidth] {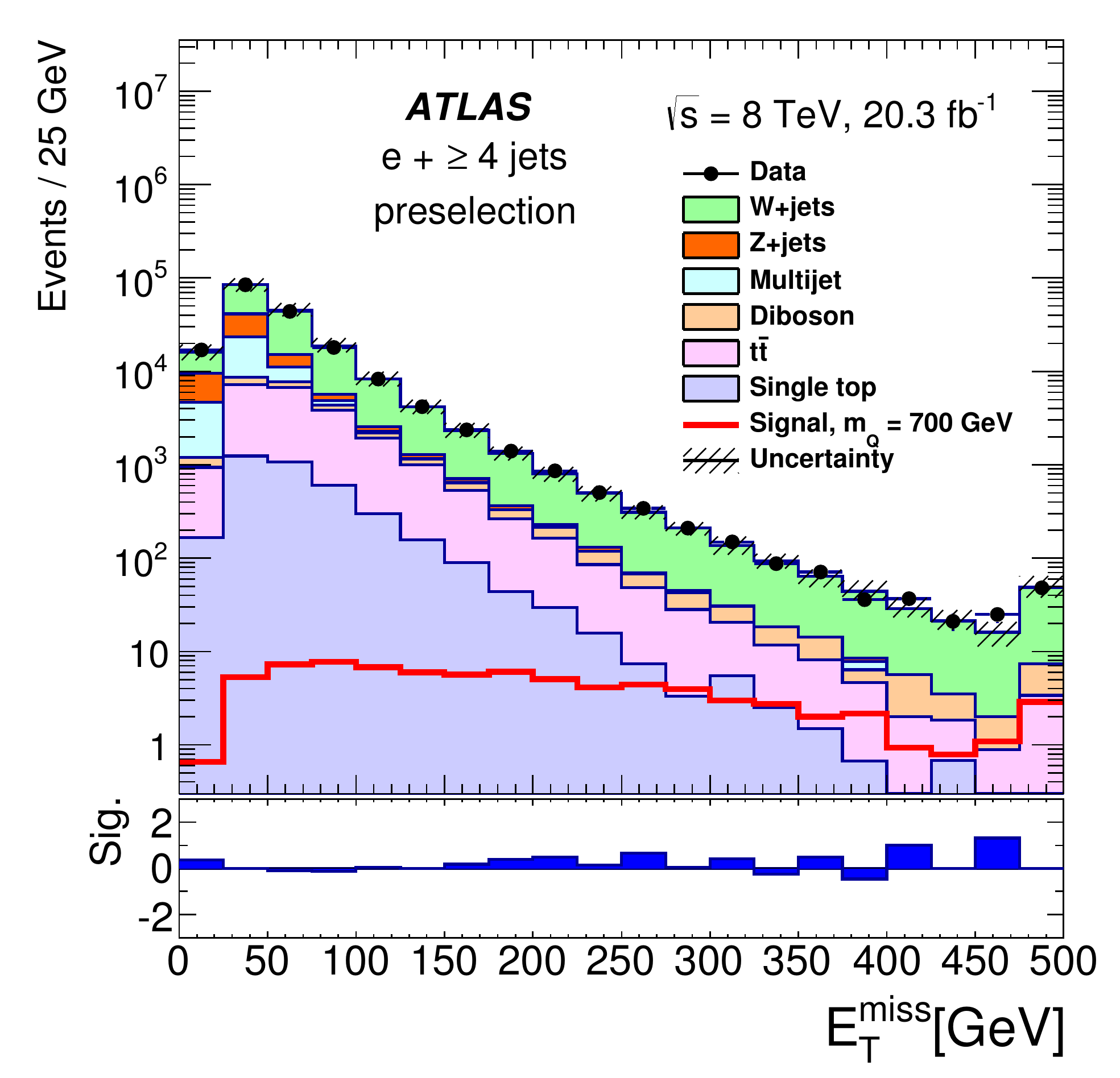}
\includegraphics*[width=0.47\textwidth] {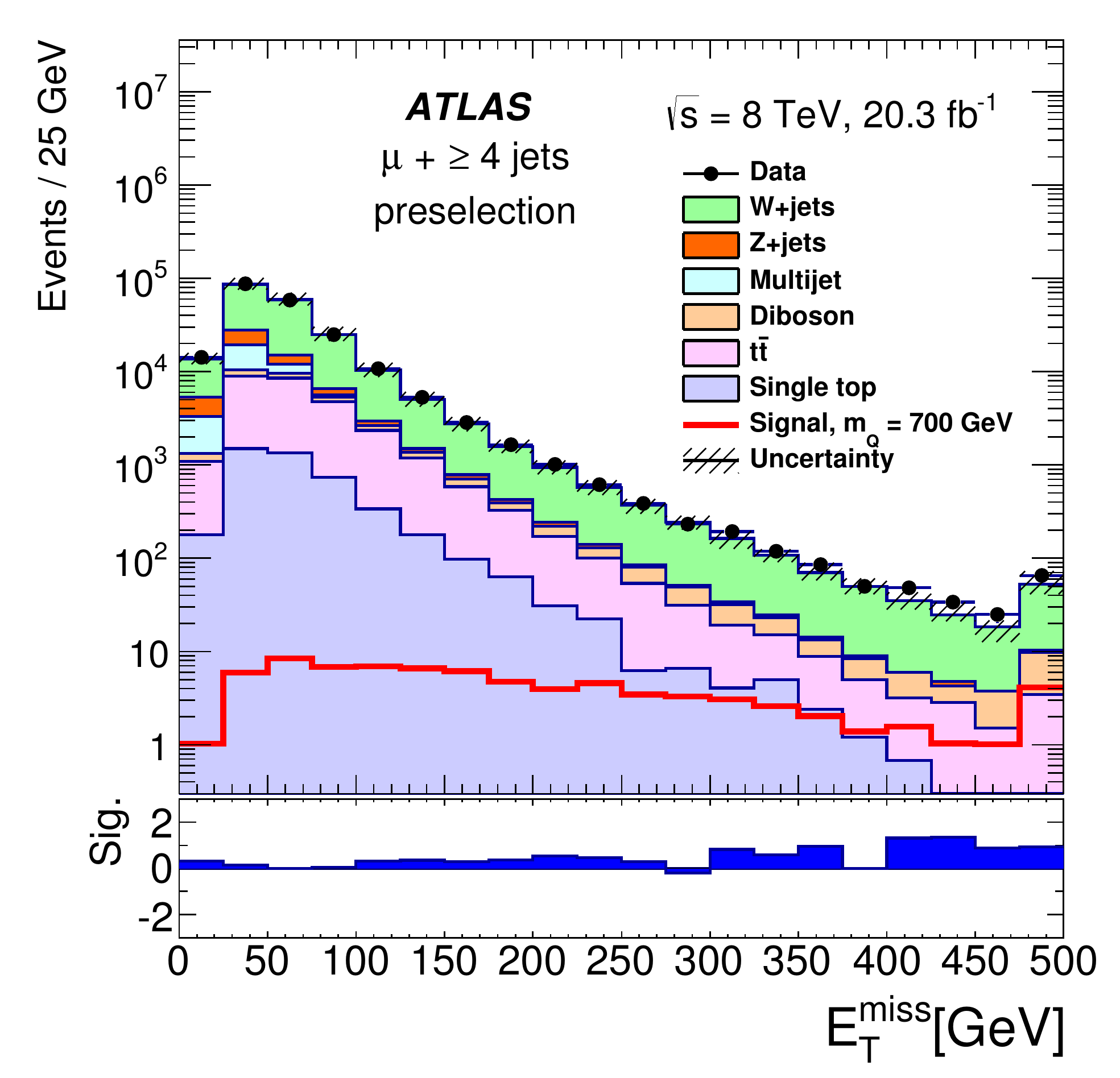}
\caption{Distributions of the missing transverse momentum after the
  $\pt^\text{truth}(V)$ correction and final normalization for the
  (left) electron and (right) muon channels.  All overflows are
  included in the rightmost bin.  The shaded band shows the total
  uncertainty on the background prediction.  The lower panel shows the
  significance of the difference between data and expectation in
  units of normal standard deviations.
\label{wptrw:fig:METstep3}
}
\end{center}
\end{figure}

\subsection{Validation regions}
\label{sec:VRcut1}

The following validation regions are used to verify the modeling of the
background processes:

\begin{itemize}
\item VR1: preselection, plus one $\Whad$ and $\Mreco< 350 \GeV$;
\item VR2: preselection, plus one $\Whad$ and $\HT < 800 \GeV$ ;
\item VR3: final selection, but with the requirements on $\pt(q_1)$,
  $\pt(q_2)$, and $\HT$ changed to $\pt(q_1) < 160\GeV$, $\pt(q_2) <
  80\GeV$, and $\HT < 800 \GeV$ and with no requirements on
  $\Delta R(\Whad, q_{1})$, $\Delta R(\Whad, q_{2})$,
  $\Delta R(\Wlep, q_{1})$, and $\Delta R(\Wlep, q_{2})$.
\end{itemize}

The validation regions are orthogonal to the signal region, but are
nevertheless useful for checking with a larger number of events that
the background normalization and kinematics are well-modeled.  The
expected signal contribution in each validation region is
approximately the size of the background uncertainty for signal masses
around the previous limit ($350\GeV$) and decreases very rapidly for
higher masses.  VR2 and VR3 are used to validate the modeling of the
final discriminant, $\Mreco$, as shown in Figs.~\ref{fig:VR2mrecocut1}
and ~\ref{fig:VR3mreco}.  VR1 is used to verify the modeling of
variables other than $\Mreco$, such as the $\HT$ distribution and the
$\pt$ spectra for the individual objects.

\begin{figure}[!h]
\begin{center}
\includegraphics*[width=0.47\textwidth] {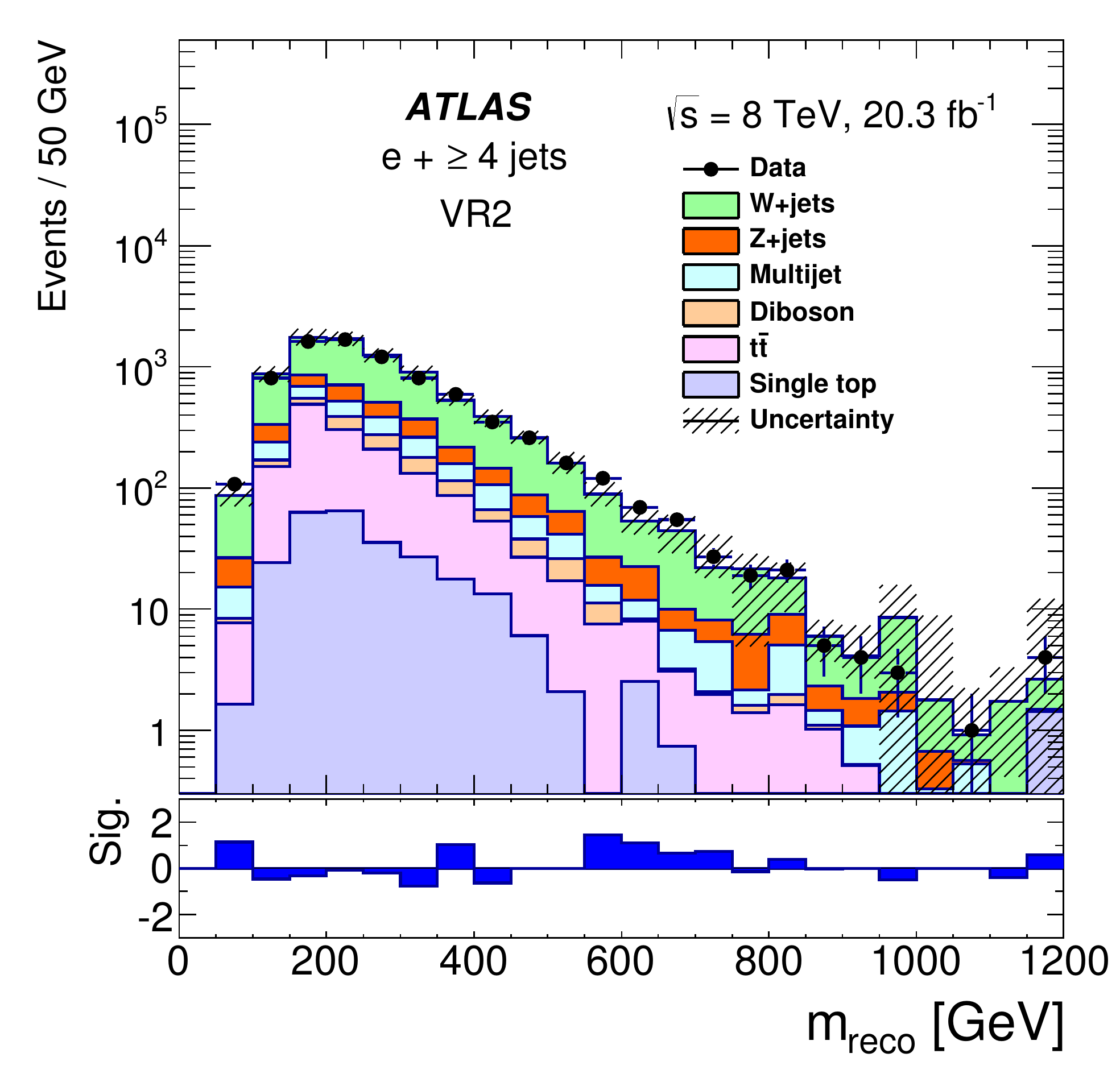}
\includegraphics*[width=0.47\textwidth] {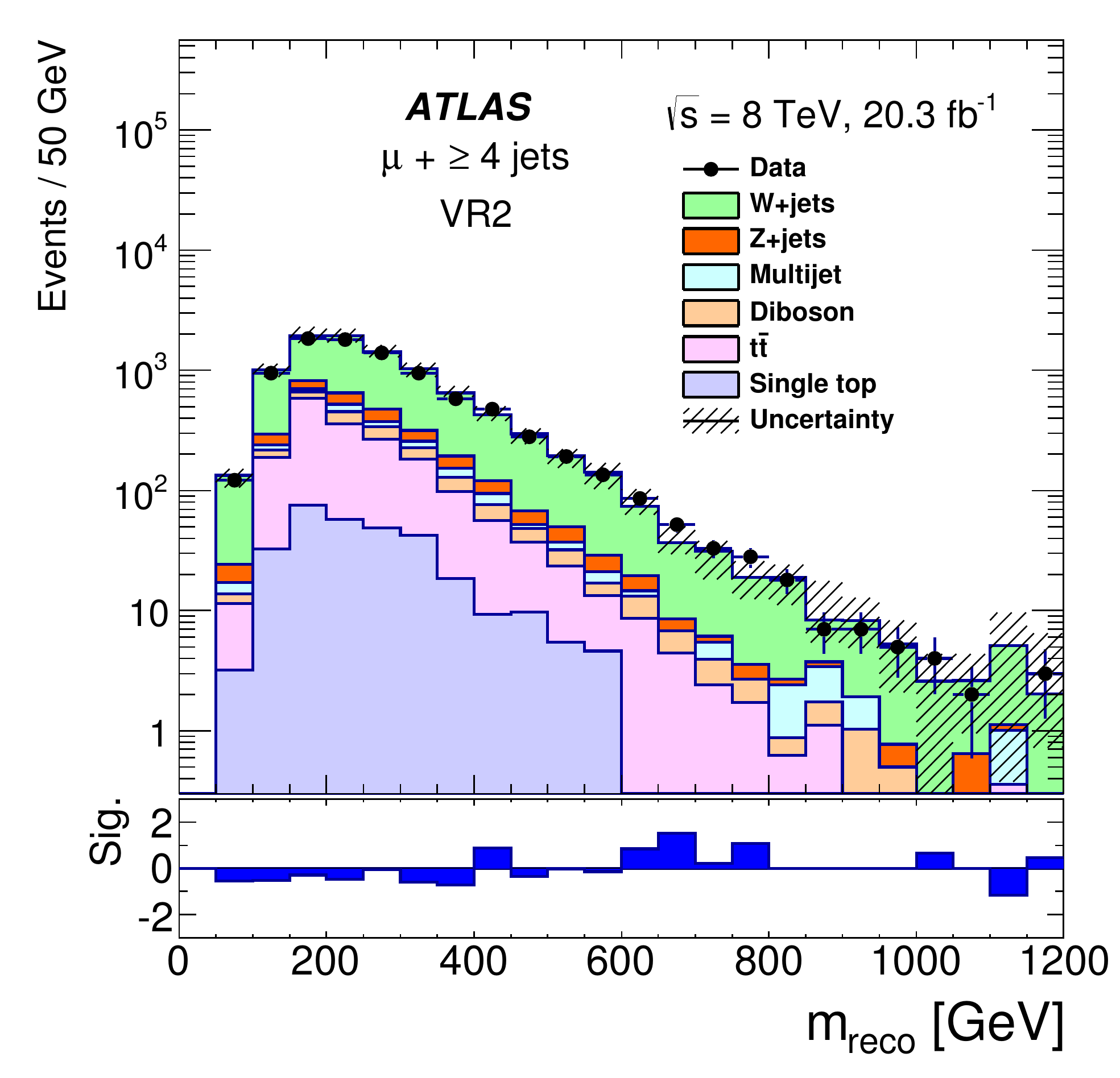}
\caption{ Comparison between data and simulation for the distribution of
$\Mreco$, the reconstructed heavy quark mass built from the
$W_{had}$ candidate and the paired $q$-jet candidate, in validation region VR2 for the (left) electron and
  (right) muon channels.  All overflows are included in the rightmost
  bin.  The shaded band shows the total uncertainty on the background
  prediction.  The lower panel shows the significance of the
  difference between data and expectation in units of normal
  standard deviations.}
\label{fig:VR2mrecocut1}
\end{center}
\end{figure}

\begin{figure}[!h]
\begin{center}
\includegraphics*[width=0.47\textwidth] {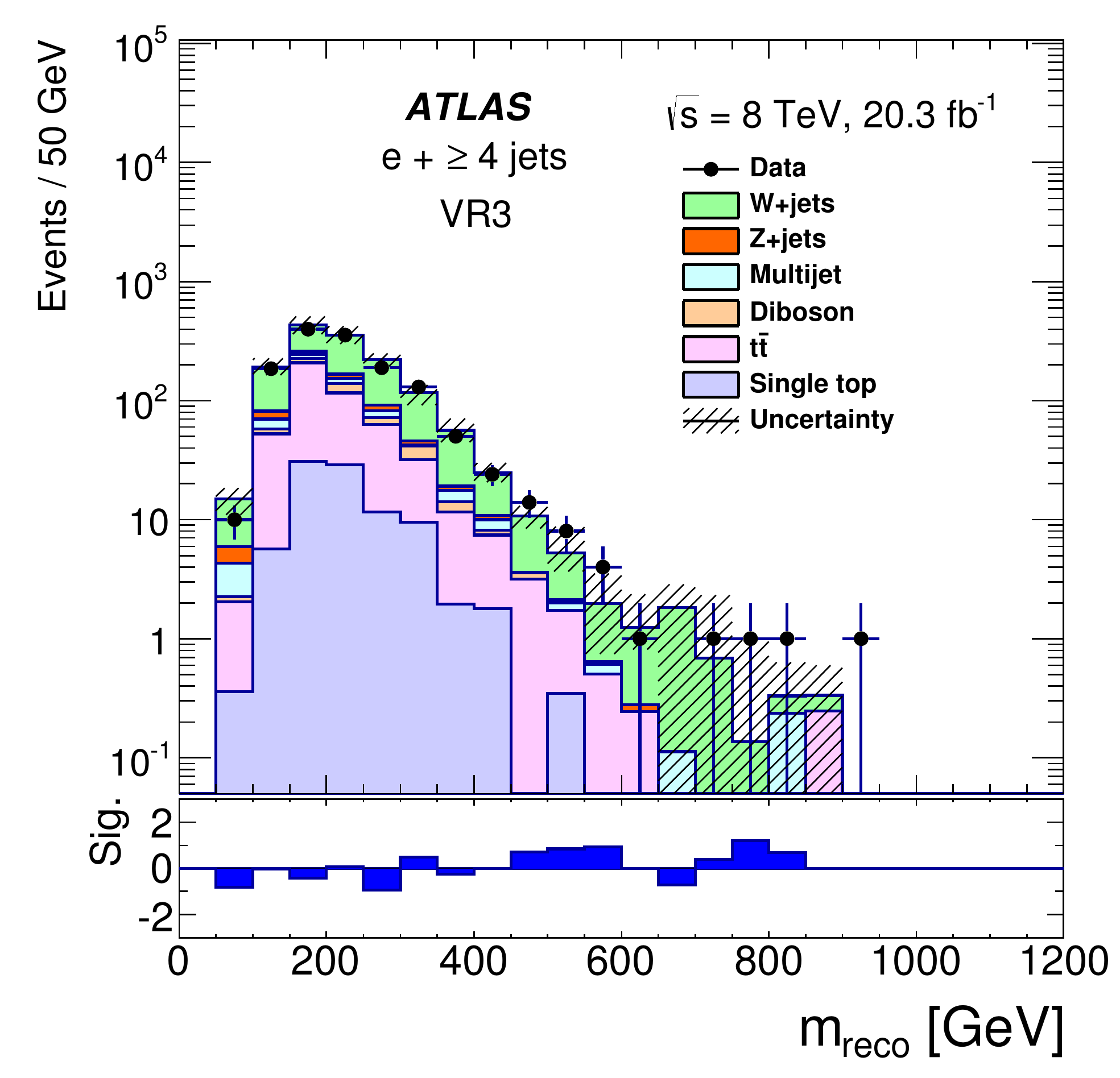}
\includegraphics*[width=0.47\textwidth] {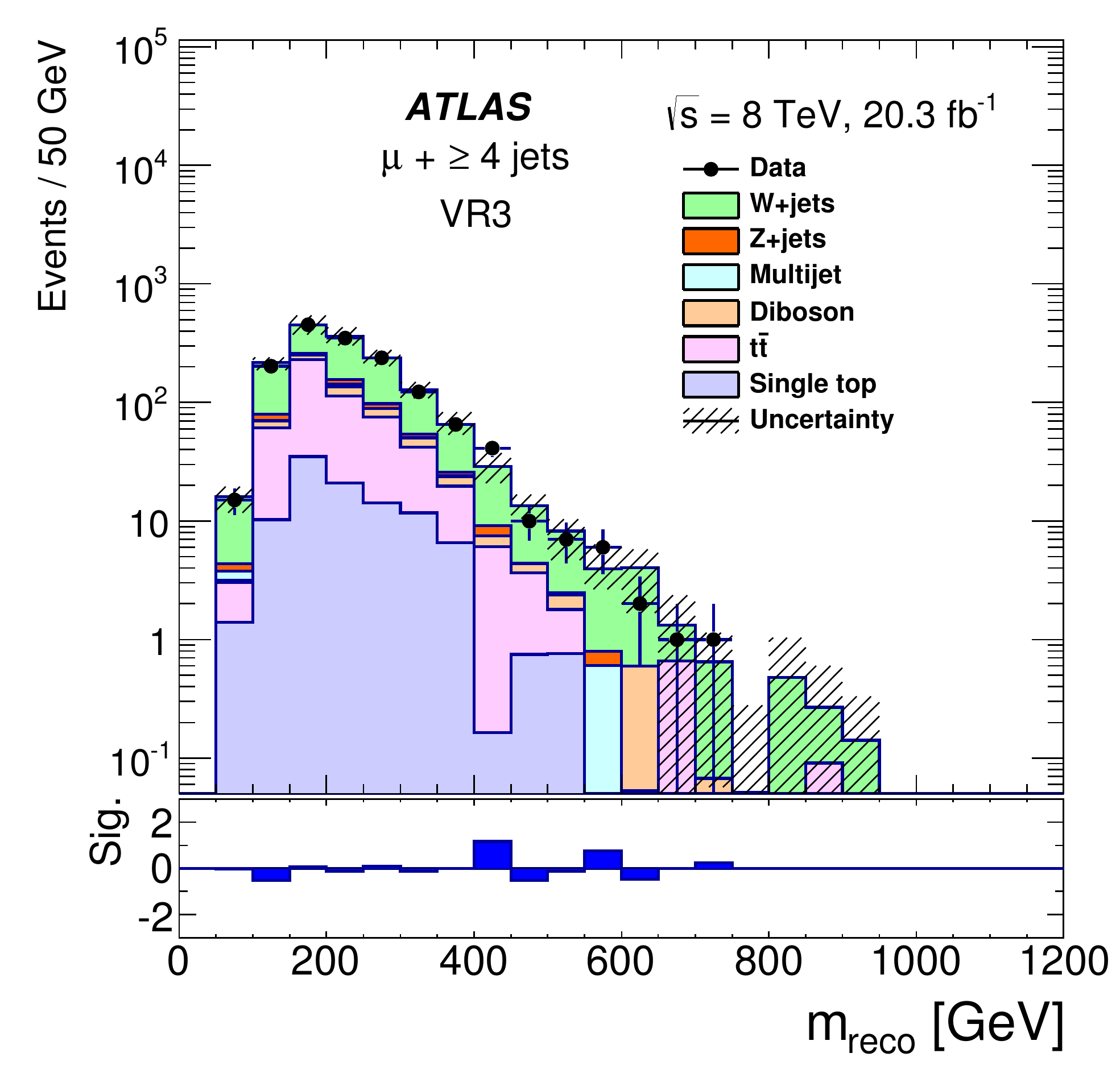}
\caption{Comparison between data and simulation for the $\Mreco$
  distribution in validation region VR3 for the (left) electron and
  (right) muon channels. The shaded band shows the total uncertainty on the background
  prediction.  The lower panel shows the significance of the
  difference between data and expectation in units of normal
  standard deviations.}
\label{fig:VR3mreco}
\end{center}
\end{figure}

\section{Systematic uncertainties}
\label {sec:systematics}
The uncertainties considered in this analysis can affect the
normalization of signal and background and the shape of the final
discriminant, \Mreco.  Each source of systematic uncertainty is
assumed to be 100\% correlated across all samples, but the different
sources are treated as uncorrelated with one another.
Table~\ref{tab:syst_table_tight} shows the impact of the dominant
uncertainties on the normalization of the background processes and a
signal sample with a mass of 700$\GeV$.

\begin{table}[h!]
\begin{center}
\renewcommand{\arraystretch}{1}
\begin{tabular}{l*{3}{c}}
\hline\hline
 & Signal $(\MQ=700\GeV)$ & Non-$W$+jets & $W$+jets\\
\hline
Luminosity & +2.8/--2.8 & +2.8/--2.8 & +2.8/--2.8\\
Normalization &  & $\pm15$ & +2.7/--4.4\\
Lepton identification & +1.6/--1.6 & +1.5/--1.5 & +1.4/--1.4\\
Jet energy resolution & +0.6/--0.6 & +12/--12 & +8.7/--8.7\\
Jet energy scale & +6.1/--4.3 & +33/--34 & +14/--18\\
$b$-tagging & +0.2/--0.2 & +5.1/--5.3 & +0.3/--0.3\\
$c$-tagging & +1.5/--1.5 & +1.5/--1.5 & +1.2/--1.2\\
Light-jet tagging & +1.0/--1.0 & +0.9/--0.9 & +1.0/--1.0\\
$\pt^\text{truth}(V)$ re-weighting &  &  & +5.7/--4.2\\

\hline
\hline
\end{tabular}

\caption{\label{tab:syst_table_tight} Table summarizing the overall
  normalization changes (expressed in \%) in signal and background
  yields for the dominant systematic uncertainties considered.  The
  selection presented here is the combination of the $e$+jets and $\mu$+jets
  channels after the final selection. }
\end{center}
\end{table}

\subsection{Normalization uncertainties}

Uncertainties affecting only the normalization include those on the
integrated luminosity ($\pm$2.8\%) and the cross sections for various
background processes. The uncertainty on the integrated luminosity is
derived following the same methodology as that detailed in
Ref.~\cite{lumi2011}.  This uncertainty is applied to all simulated
signal and background processes.

After the final selection, the non-$W$+jets background has a
total normalization uncertainty of 15\%.  The predicted contribution
from the multijet background is negligible compared to the uncertainty
on the non-$W$+jets background, so it is neglected.

As described in Sec.~\ref{sec:wjets_model}, the normalization of the
$W$+jets background is determined from a fit to data using both lepton
channels.  The uncertainty on the $W$+jets normalization is determined
by comparing that result to the normalization one would obtain if only
the electron or only the muon channel were used.  This is motivated by
the fact that it is not known whether the normalization from the
electron channel or the muon channel (or something in between) is
correct and this procedure leads to an uncertainty of +2.7/--4.4\%.
The statistical uncertainty from the fit is negligible (0.03\%) in
comparison.

The rest of the systematic uncertainties can modify both the
normalization and shape of the $\Mreco$ distribution.

\subsection{Shape uncertainties}

Uncertainties on the trigger, reconstruction, and isolation
efficiencies for the selected lepton are estimated using $Z\to ee$ and
$Z\to \mu \mu$ events~\cite{Aad:2014rra, Aad:2014nim}. In addition,
high jet-multiplicity $Z\to \ell\ell$ events are studied, from which
extra uncertainties on the isolation efficiency are assigned to
account for the difference between $Z$ boson and $t\bar{t}$
events. Uncertainties on the $E^{\rm miss}_\mathrm{T}$ reconstruction
and the energy scale and resolution of the leptons were also
considered; however, these have a very small impact on the results.

The jet energy resolution is measured by studying dijet events in
data and simulation. The simulation is found to agree with data to
better than 10\%~\cite{Aad:2012ag}. The differences in resolutions
between data and simulations are used to determine the relative
systematic uncertainty.  The uncertainty on the jet energy scale is
evaluated by repeating the analysis with the jet energy scale shifted
by $\pm1 \sigma$~\cite{Aad:2011he, Aad:2014bia}.  The jet
reconstruction efficiency is estimated using track-based jets and is
well described in simulation.  To account for differences in the
efficiency for reconstructing jets in simulated events compared to
collider data, the efficiencies are measured in both samples and the
difference is taken as the uncertainty.
The uncertainty due to the JVF requirement is evaluated by comparing
the signal and background distributions with the JVF cut shifted up
and down by 10\%, a variation that spans the difference observed
between data and simulation in this quantity.
The $b$-tagging efficiency for $b$-jets, as well as $c$-jets and
light jets, is derived in data and a simulated $t\bar{t}$ sample,
parameterized as a function of $\pt$ and $\eta$~\cite{btagConf,
  bcltag}. The corresponding (in)efficiencies in the simulated samples
are corrected to match those in data and the uncertainties from the
calibration are propagated through the analysis.

The $W$+jets sample is assigned a $\pt^\text{truth}(V)$-dependent
shape uncertainty due to the correction described in
Sec.~\ref{sec:wjets_model}.  Four sources of uncertainty on the
$\pt^\text{truth}(V)$ correction are considered: (1) the statistical
uncertainty on the parameters of the reweighting function; (2) the
difference between the nominal reweighting function and alternative
corrections obtained by only considering the electron or muon channel;
(3) the dependence of the fit on the choice of bin width when deriving
the reweighting function; (4) the difference between alternative
parameterizations of the reweighting function.
A closure test is performed to verify that any residual differences
between data and prediction are well within the assigned uncertainty.

\section{Results}
\label {sec: results}

The final $\Mreco$ distribution for the combined electron and muon
channels is shown in Fig.~\ref{fig:inputslimits} for three signal
scenarios: $m_Q = 600\GeV$, $m_Q = 700\GeV$, and $m_Q = 800\GeV$.  The
observed distribution shows a slight excess over the SM expectation,
but the excess is broader than expected for signal and is consistent
with the background-only prediction at the level of two standard
deviations.  Therefore, the analysis proceeds to setting limits on the
signal hypothesis.

The $\Mreco$ distribution for the combined electron and muon channels
after the final selection (Fig.~\ref{fig:inputslimits}) is used to
derive 95\% confidence level (CL) limits on the $Q\bar{Q}$ production
cross section using the $CL_{s}$ method~\cite{Junk:1999kv,
  Read:2002hq}.  

\begin{figure}[htb]
  \begin{center}
    \includegraphics[width=0.48\textwidth]{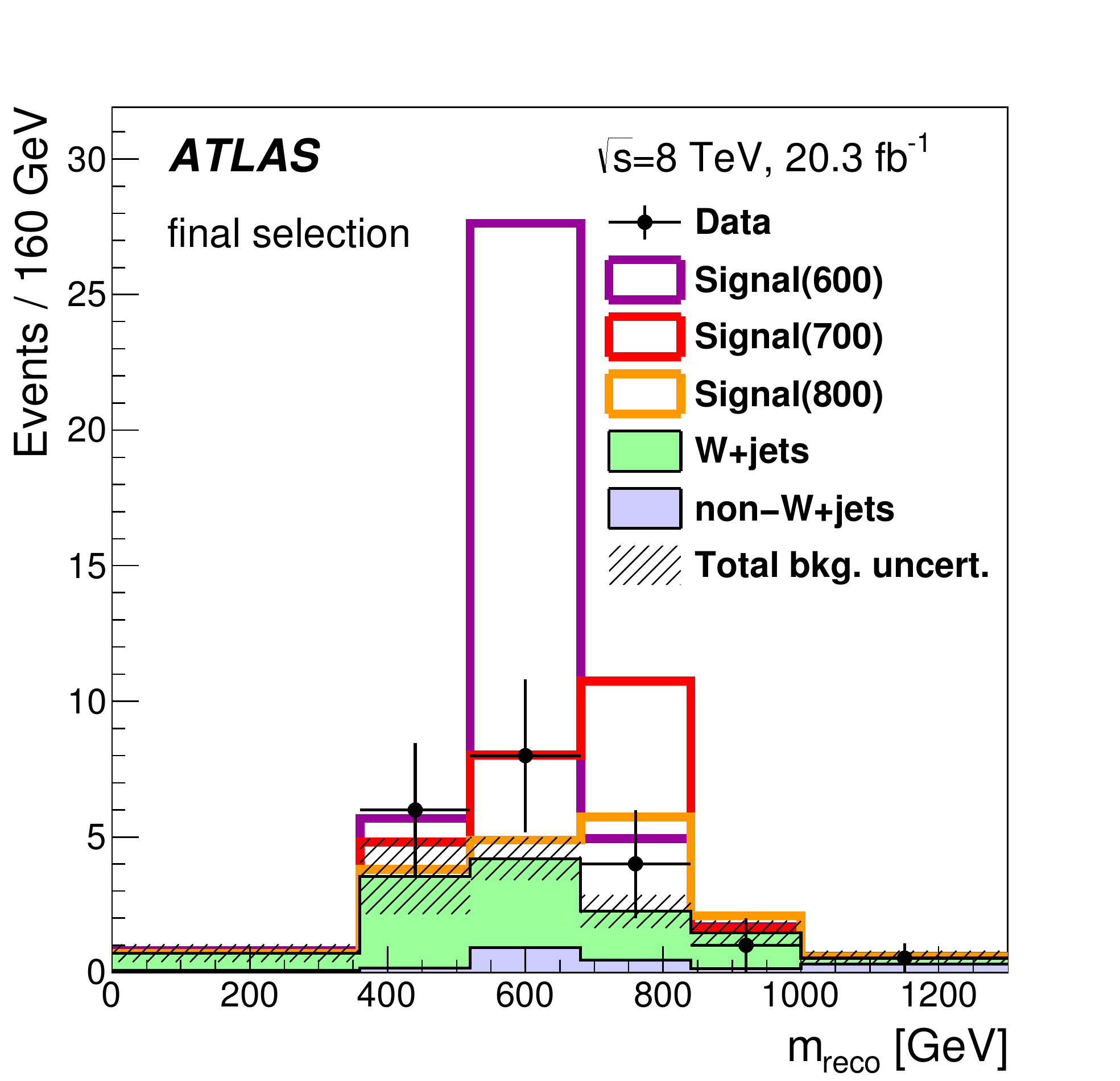}
 \caption{Distribution of the final discriminant, the mass of the
   hadronically decaying heavy quark candidate, with the bin widths
   chosen as for the statistical analysis. The expected contribution
   from signal with BR($Q\to Wq$)=1 is shown stacked on top of the
   total background prediction for three mass scenarios, $m_Q = 600,
   700,$ and $800\GeV$.  The backgrounds are stacked with the largest
   on top. The shaded band shows the total uncertainty on the
   background prediction.  }
    \label{fig:inputslimits}
  \end{center}
\end{figure}

Limits on the pair production of new chiral quarks are evaluated by
setting BR$(Q\to Wq)=1$.  Figure~\ref{fig:exclusionplots} shows the
observed and expected limits on a heavy chiral quark as a function of
$\MQ$, compared to the theoretical
prediction~\cite{Baernreuther:2012ws, Czakon:2012zr, Czakon:2012pz,
  Czakon:2013goa, Cacciari:2011hy, Czakon:2011xx}.  The total
uncertainty on the theoretical cross section includes the
contributions from the scale variations and PDF uncertainties.  Using
the central value of the theoretical cross section, the observed
(expected) 95\% CL limit on the mass of a new chiral quark is $\MQ >
690\GeV$ $(780\GeV$).
This represents the most stringent limit to date on the mass of a new
quark decaying exclusively into a $W$ boson and a light quark ($u, d,
s$).  This limit is also applicable to the production of pairs of
down-type vectorlike quarks with electric charge of $-4/3$ which each
decay into a $W$ boson and a light quark ($d, s$).

\begin{figure}[h!]
\begin{center}
\begin{tabular}{cc}
\includegraphics[width=0.5\textwidth]{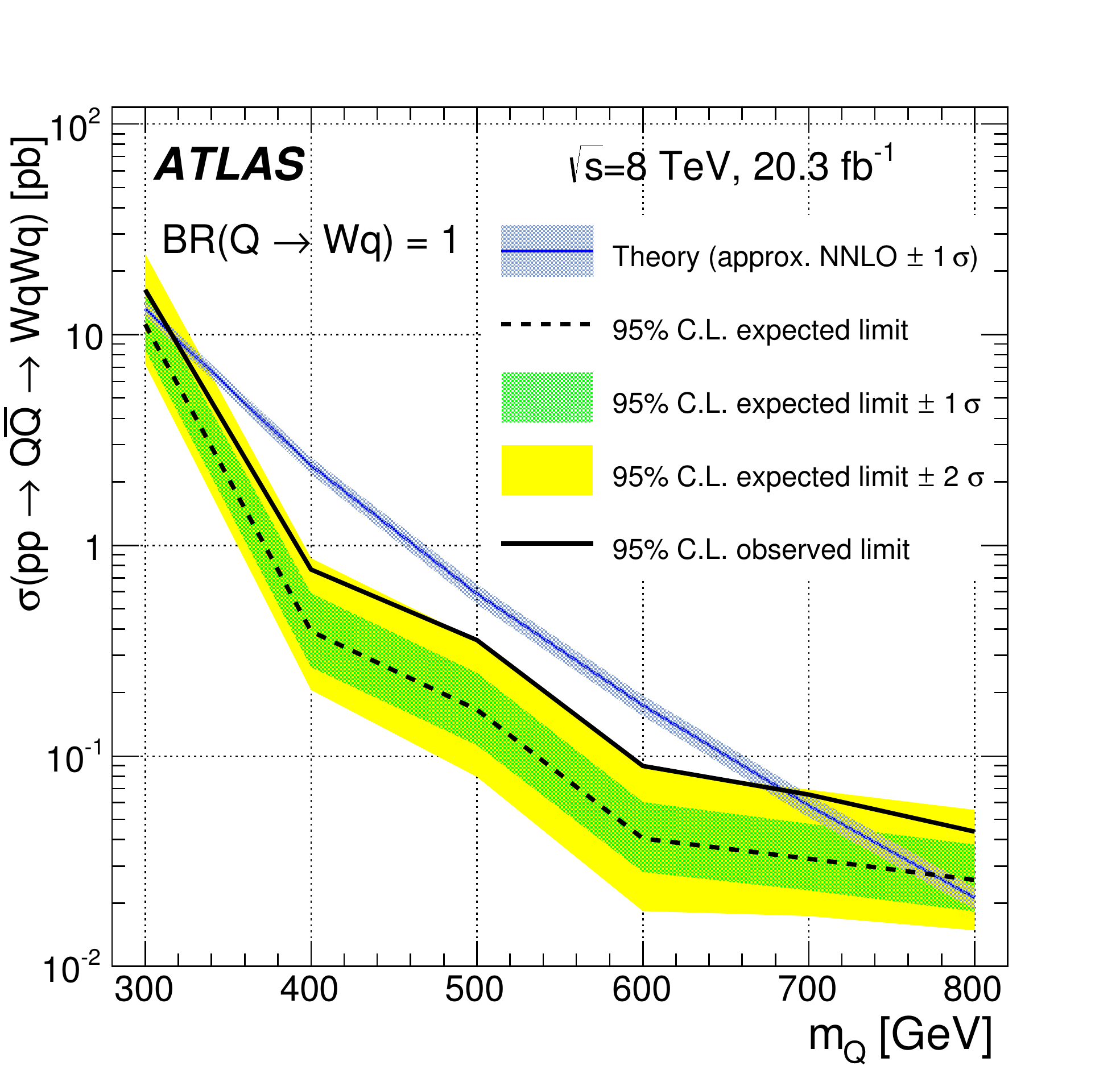}
\end{tabular}
\caption{The expected (black dashed line) and observed (black solid
  line) 95\% CL upper limits on the cross section as a function of
  $\MQ$ when setting BR$(Q\to Wq) = 1$, which would be the case for a
  new chiral quark.  The green and yellow shaded bands indicate the
  $\pm 1\sigma$ and $\pm 2\sigma$ intervals on the distribution
  of expected results for the chiral model if no signal exists.  The
  blue line is the theoretical prediction for the total cross section
  [i.e., assuming BR$(Q\to Wq)=1$] with its uncertainties.}
\label{fig:exclusionplots}
\end{center}
\end{figure}

Next, the VLQ signal samples are used to set limits on the mass of a
heavy quark that decays to a light quark ($u, d, s$) and either a $W$,
$Z$, or $H$ boson.  The results are given as a function of the
branching ratios BR$(Q\to Wq)$ versus BR$(Q\to Hq)$, with the
branching ratio to $Zq$ fixed by the requirement
BR$(Q\!\to\!Zq) = 1 - \text{BR}(Q\!\to\!Wq) - \text{BR}(Q\!\to\!Hq)$.
The analysis loses sensitivity at low masses due to the tight
selection requirements optimized for the decay $Q\to Wq$, so the
results are presented as the upper and lower bounds on the mass range
that is excluded at 95\% CL.  The expected limits on $\MQ$ as a
function of the branching ratios are shown in Fig.~\ref{fig:exp2d} and
the observed limits are shown in Fig.~\ref{fig:obs2d}.  For example,
for the branching ratios BR$(Q\!\to\!Wq) = 0.6$, BR$(Q\!\to\!Zq) =
0.4$, and BR$(Q\!\to\!Hq) = 0$, the results exclude VLQs with a mass
from $370\GeV$ to $610\GeV$.  If no signal is present, the median
expected exclusion range for that set of branching ratios is $340\GeV$
to $690\GeV$.
For BR$(Q\!\to\!Wq) = 0.6$, BR$(Q\!\to\!Zq) = 0.3$, and
BR$(Q\!\to\!Hq) = 0.1$, the results exclude VLQs with a mass from
$370\GeV$ to $470\GeV$, while the median expected exclusion range is
$340\GeV$ to $680\GeV$.
For BR$(Q\!\to\!Wq) = 0.5$, BR$(Q\!\to\!Zq) = 0.25$, and
BR$(Q\!\to\!Hq) = 0.25$, the results exclude VLQs with a mass from
$390\GeV$ to $410\GeV$, while the median expected exclusion range is
$360\GeV$ to $650\GeV$.

\begin{figure}[h!]
\begin{center}
\begin{tabular}{cc}
\includegraphics[width=0.5\textwidth]{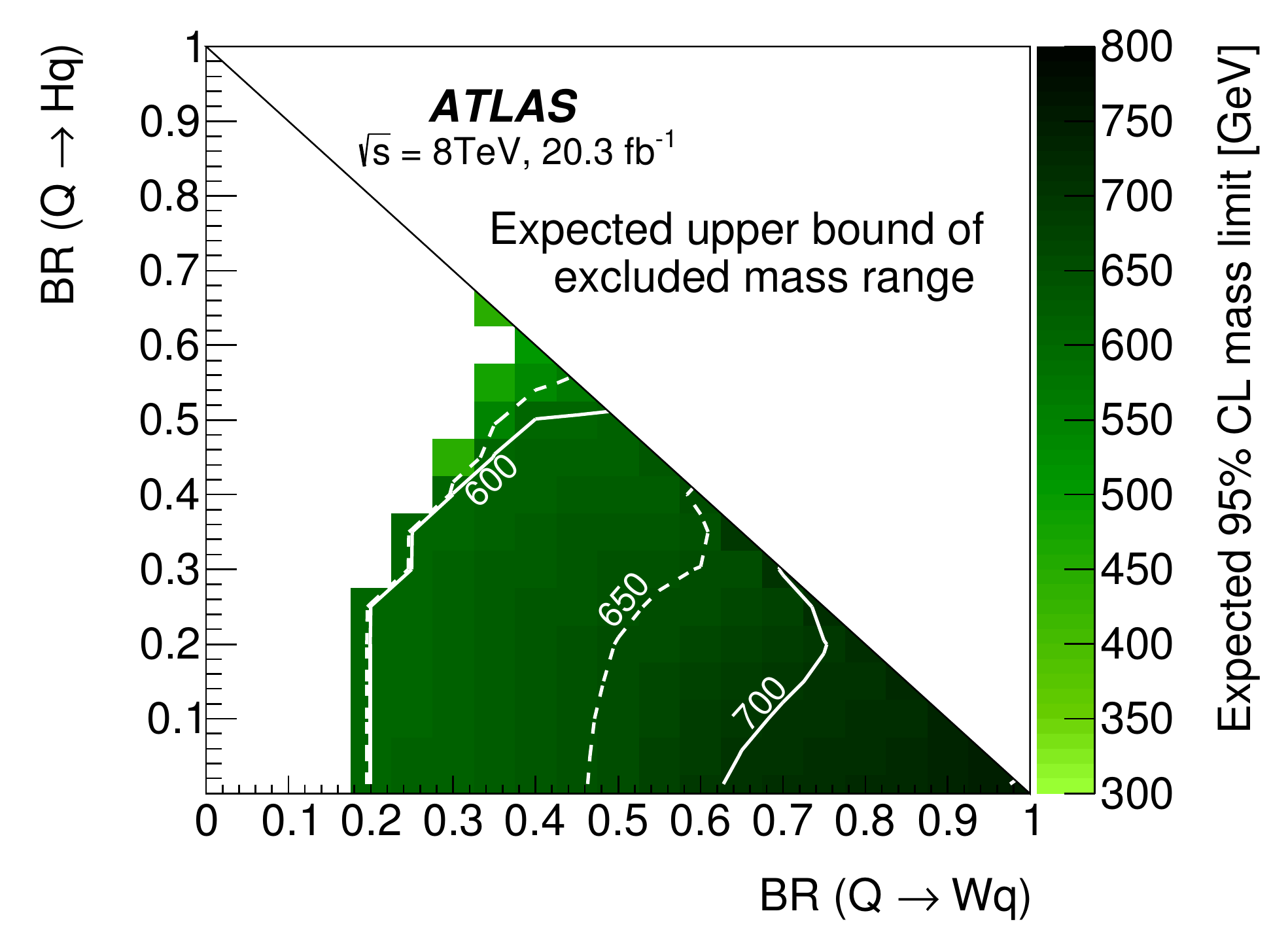}
\includegraphics[width=0.5\textwidth]{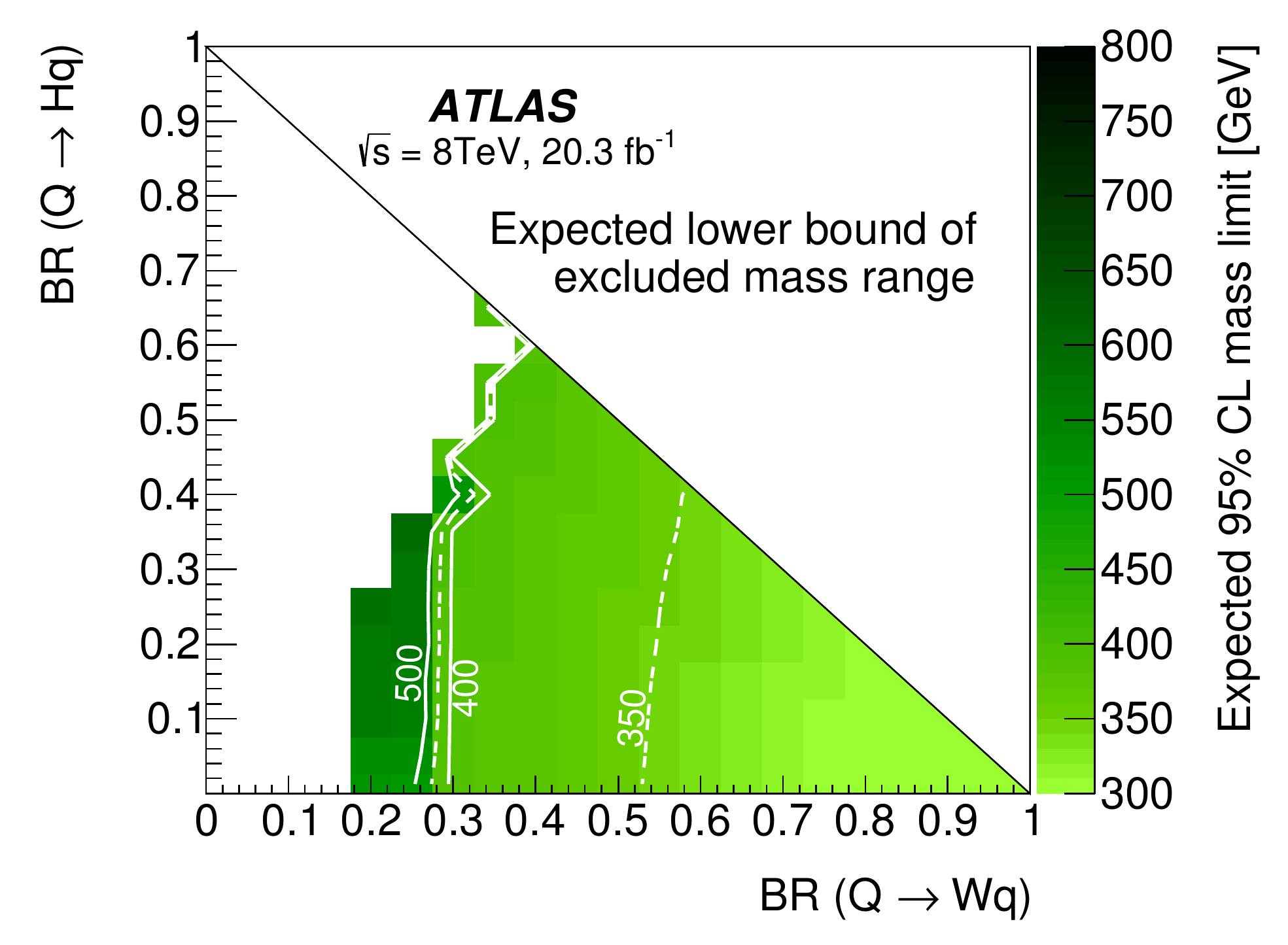}
\end{tabular}
\caption{The (left) upper and (right) lower bounds on the range of
  heavy quark masses expected to be excluded at 95\% CL, as a function
  of the branching ratio of the heavy quark to $Wq$ versus $Hq$, with the
branching ratio to $Zq$ fixed by the requirement
BR$(Q\!\to\!Zq) = 1 - \text{BR}(Q\!\to\!Wq) - \text{BR}(Q\!\to\!Hq)$.
  The region above the diagonal is forbidden by unitarity.}
\label{fig:exp2d}
\end{center}
\end{figure}

\begin{figure}[h!]
\begin{center}
\begin{tabular}{cc}
\includegraphics[width=0.5\textwidth]{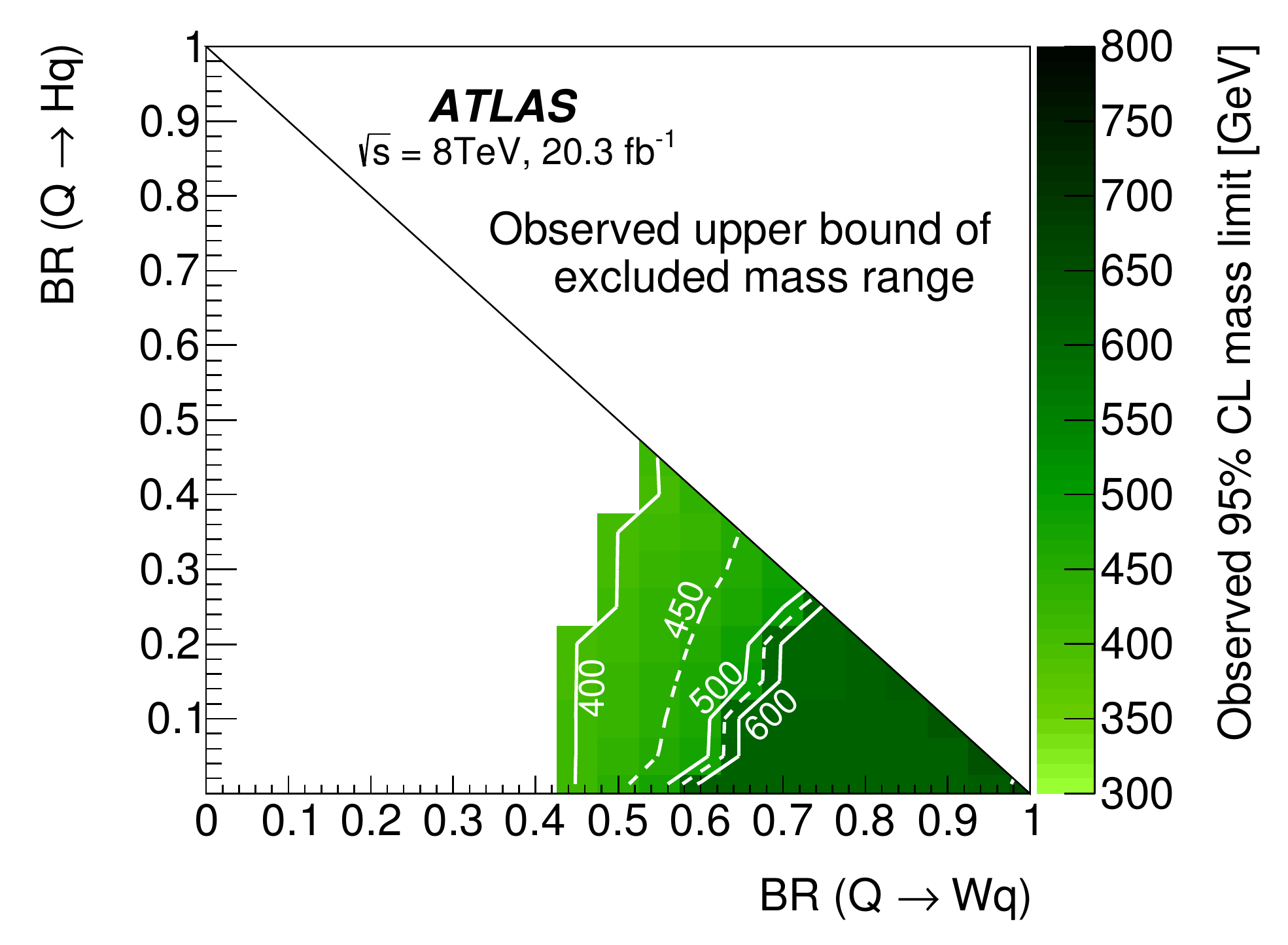}
\includegraphics[width=0.5\textwidth]{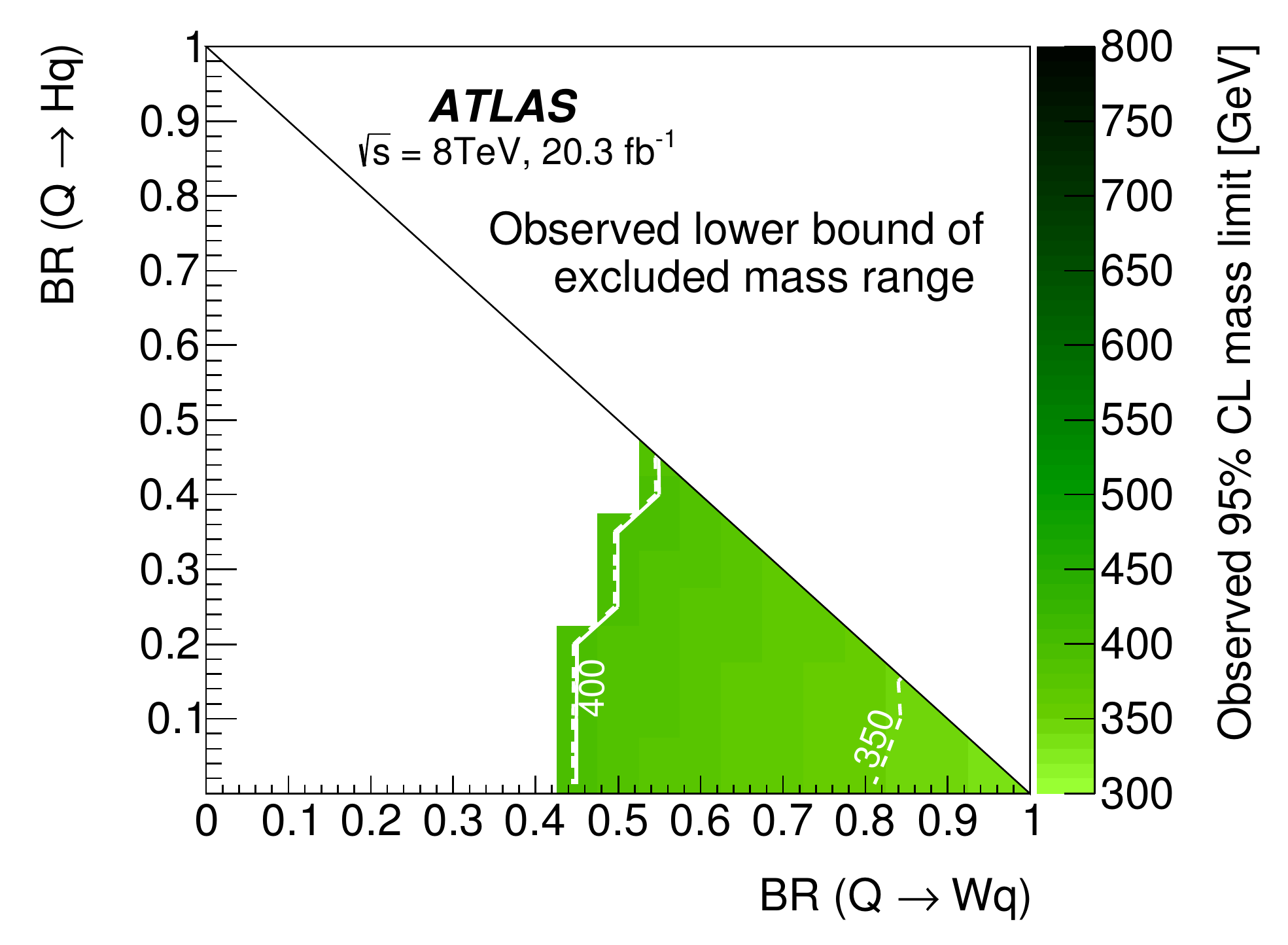}
\end{tabular}
\caption{The (left) upper and (right) lower bounds on the range of
  heavy quark masses observed to be excluded at 95\% CL, as a function
  of the branching ratio of the heavy quark to $Wq$ versus $Hq$, with the
branching ratio to $Zq$ fixed by the requirement
BR$(Q\!\to\!Zq) = 1 - \text{BR}(Q\!\to\!Wq) - \text{BR}(Q\!\to\!Hq)$.
  The region above the diagonal is forbidden by unitarity.}
\label{fig:obs2d}
\end{center}
\end{figure}

\FloatBarrier

\section{Conclusion}
\label{sec:conclusion} 
A search for new heavy quarks that decay to a $W$ boson and a light
quark $(u, d, s)$ is performed with a dataset corresponding to
20.3\,\ifb{} that was collected by the ATLAS detector at the LHC in
$pp$ collisions at $\sqrt{s} = 8\TeV$.  No significant evidence of a
heavy quark signal is observed when selecting events with one charged
lepton, large $\met$, and four non-$b$-tagged jets.  The results
exclude new heavy chiral quarks with masses up to $690\GeV$ at 95\%
CL. This search is also interpreted in the context of vectorlike quark
models, for which the new heavy quark can decay to a light quark and
either a $W$, $Z$, or $H$ boson.  New exclusions limits are presented
in the two-dimensional plane of BR$(Q\to Wq)$ versus BR$(Q\to Hq)$.

\input{Acknowledgements}

\printbibliography

\newpage
\input{atlas_authlist}

\end{document}

%% file: Acknowledgements.tex

\section{Acknowledgements}
We thank CERN for the very successful operation of the LHC, as well as the
support staff from our institutions without whom ATLAS could not be
operated efficiently.

We acknowledge the support of ANPCyT, Argentina; YerPhI, Armenia; ARC,
Australia; BMWFW and FWF, Austria; ANAS, Azerbaijan; SSTC, Belarus; CNPq and FAPESP,
Brazil; NSERC, NRC and CFI, Canada; CERN; CONICYT, Chile; CAS, MOST and NSFC,
China; COLCIENCIAS, Colombia; MSMT CR, MPO CR and VSC CR, Czech Republic;
DNRF, DNSRC and Lundbeck Foundation, Denmark; EPLANET, ERC and NSRF, European Union;
IN2P3-CNRS, CEA-DSM/IRFU, France; GNSF, Georgia; BMBF, DFG, HGF, MPG and AvH
Foundation, Germany; GSRT and NSRF, Greece; RGC, Hong Kong SAR, China; ISF, MINERVA, GIF, I-CORE and Benoziyo Center, Israel; INFN, Italy; MEXT and JSPS, Japan; CNRST, Morocco; FOM and NWO, Netherlands; BRF and RCN, Norway; MNiSW and NCN, Poland; GRICES and FCT, Portugal; MNE/IFA, Romania; MES of Russia and NRC KI, Russian Federation; JINR; MSTD,
Serbia; MSSR, Slovakia; ARRS and MIZ\v{S}, Slovenia; DST/NRF, South Africa;
MINECO, Spain; SRC and Wallenberg Foundation, Sweden; SER, SNSF and Cantons of
Bern and Geneva, Switzerland; NSC, Taiwan; TAEK, Turkey; STFC, the Royal
Society and Leverhulme Trust, United Kingdom; DOE and NSF, United States of
America.

The crucial computing support from all WLCG partners is acknowledged
gratefully, in particular from CERN and the ATLAS Tier-1 facilities at
TRIUMF (Canada), NDGF (Denmark, Norway, Sweden), CC-IN2P3 (France),
KIT/GridKA (Germany), INFN-CNAF (Italy), NL-T1 (Netherlands), PIC (Spain),
ASGC (Taiwan), RAL (UK) and BNL (USA) and in the Tier-2 facilities
worldwide.

%% file: atlas_authlist.tex
\begin{flushleft}
{\Large The ATLAS Collaboration}

\bigskip

G.~Aad$^{\rm 85}$,
B.~Abbott$^{\rm 113}$,
J.~Abdallah$^{\rm 151}$,
O.~Abdinov$^{\rm 11}$,
R.~Aben$^{\rm 107}$,
M.~Abolins$^{\rm 90}$,
O.S.~AbouZeid$^{\rm 158}$,
H.~Abramowicz$^{\rm 153}$,
H.~Abreu$^{\rm 152}$,
R.~Abreu$^{\rm 116}$
% ...
\bigskip
\\
% ...
$^{11}$ Institute of Physics, Azerbaijan Academy of Sciences, Baku, Azerbaijan\\
% ...
$^{85}$ CPPM, Aix-Marseille Universit{\'e} and CNRS/IN2P3, Marseille, France\\
% ...
$^{90}$ Department of Physics and Astronomy, Michigan State University, East Lansing MI, United States of America\\
% ...
$^{107}$ Nikhef National Institute for Subatomic Physics and University of Amsterdam, Amsterdam, Netherlands\\
% ...
$^{113}$ Homer L. Dodge Department of Physics and Astronomy, University of Oklahoma, Norman OK, United States of America\\
% ...
$^{116}$ Center for High Energy Physics, University of Oregon, Eugene OR, United States of America\\
% ...
$^{151}$ Institute of Physics, Academia Sinica, Taipei, Taiwan\\
$^{152}$ Department of Physics, Technion: Israel Institute of Technology, Haifa, Israel\\
$^{153}$ Raymond and Beverly Sackler School of Physics and Astronomy, Tel Aviv University, Tel Aviv, Israel\\
% ...
$^{158}$ Department of Physics, University of Toronto, Toronto ON, Canada\\
% ...
\end{flushleft}
